\documentclass{svjour3}                     
\smartqed  
\usepackage{amssymb}
\usepackage{graphicx}
\usepackage{natbib}
\usepackage{mathptmx}
\bibpunct{(}{)}{;}{a}{}{,}
\newcommand{\be}{\begin{equation}}
\newcommand{\ee}{\end{equation}}
\newcommand{\beq}{\begin{eqnarray}}
\newcommand{\eeq}{\end{eqnarray}}
\newcommand\subsun[1]{{$_{\normalsize\odot}$}}
\newcommand{\adv}[3]{   #1, { Adv. Space Res.},    {  #2}, #3}
\newcommand{\aeta}[3]{  #1, { A\&A}, {  #2}, #3}
\newcommand{\aetap}[1]{ #1, { Astron. Astrophys.}, in press}

\newcommand{\araa}[3]{  #1, { ARA\&A}, {  #2}, #3}
\newcommand{\aj}[3]{  #1, { AJ}, {  #2}, #3}
\newcommand{\aspj}[3]{  #1, { ApJ}, {  #2}, #3}

\newcommand{\mnras}[3]{   #1, { MNRAS}, {  #2}, #3}

\newcommand{\SpaceS}[3]{#1, { Space Sci. Rev.}, {  #2}, #3}

 \journalname{SSRv}

\begin{document}

\title{Observations of extended radio emission in clusters}

\titlerunning{Extended radio emission in clusters}

\author{C.~Ferrari \and
        F.~Govoni \and
        S.~Schindler \and        
        A.M.~Bykov \and 
        Y.~Rephaeli}

\authorrunning{Ferrari et al.}

\institute{C. Ferrari \at Institute for Astro- and Particle Physics, 
                 University of Innsbruck, Technikerstr. 25, 6020
                 Innsbruck, Austria \\
                  \email{chiara.ferrari@uibk.ac.at}
           \and
           F. Govoni \at INAF - Osservatorio Astronomico di Cagliari,
           Loc. Poggio dei Pini, Strada 54, 09012 Capoterra, Cagliari,
           Italy 
           \and
           S. Schindler \at Institute for Astro- and Particle Physics, 
                 University of Innsbruck, Technikerstr. 25, 6020 
                 Innsbruck, Austria 
           \and
           A.M. Bykov \at A.F. Ioffe Institute for Physics and Technology, 194021 
                 St. Petersburg, Russia
           \and
           Y. Rephaeli \at School of Physics \& Astronomy, Tel Aviv University, 
            Tel Aviv, 69978, Israel
             }

\date{Received: 5 October 2007 ; Accepted: 11 November 2007 }

\maketitle

\begin{abstract}
We review observations of extended regions of radio emission in
clusters; these include diffuse emission in `relics', and the large
central regions commonly referred to as `halos'. The spectral
observations, as well as Faraday rotation measurements of background
and cluster radio sources, provide the main evidence for large-scale
intracluster magnetic fields and significant densities of relativistic
electrons. Implications from these observations on acceleration
mechanisms of these electrons are reviewed, including turbulent and
shock acceleration, and also the origin of some of the electrons in
collisions of relativistic protons by ambient protons in the (thermal)
gas. Improved knowledge of non-thermal phenomena in clusters requires
more extensive and detailed radio measurements; we briefly review
prospects for future observations.  
\end{abstract}

\section{Introduction}
\label{Introduction} 

In the last 10 years, the improved capabilities (sensitivity, spectral
and spatial resolution) of multi-wavelength telescopes have allowed us
to study in detail the formation and evolution of the largest
gravitationally bound systems in the Universe, i.e. galaxy
clusters. Following the hierarchical scenario of structure formation,
massive clusters form through episodic mergers of smaller mass units
(groups and poor clusters) and through the continuous accretion of
field galaxies.

It has now been proven that major cluster mergers, with their huge
release of gravitational binding energy ($\sim 10^{64}$ ergs), deeply
affect the physical properties of the different components of
clusters, i.e. the temperature, metallicity and density distribution
of the thermal intracluster medium (ICM) emitting in X-rays
\citep[e.g.][]{Buote02, Sauvageot05, Ferrari06a, Kapferer06,
  Markevitch07}, the global dynamics and spatial distribution of
galaxies \citep[e.g.][]{Girardi02, Ferrari03b, Ferrari05,
  Maurogordato07}, as well as their star-formation rate
\citep[e.g.][]{Gavazzi03, Poggianti04, Ferrari06b}. The typical signatures
that allow to identify merging clusters from optical and X-ray
observations are: a) substructures in the X-ray and optical surface
densities \citep[see][and references therein]{Buote02}, b)
non-Gaussian radial velocity distributions of cluster members
\citep[see][and references therein]{Girardi02}, c) patchy ICM
temperature, pressure, entropy and metallicity maps
\citep[e.g.][]{Finoguenov05, Kapferer06}, d) sharp X-ray surface
brightness discontinuities, accompanied by jumps in gas and
temperature \citep[``cold fronts'', see, e.g.,][]{Markevitch00}, e) absent or
disturbed cooling-cores \citep[e.g.][]{Markevitch99}, f) larger core
radii compared to (nearly) relaxed clusters
\citep[][]{Jones99}. There are also indications that recent
merging events lead to a depletion of the nearest cluster neighbours
\citep[e.g.][]{Schuecker99}. Additionally, deep radio observations
have revealed the presence of diffuse and extended ($\sim$ 1 Mpc)
radio sources in about 50 merging clusters. Their radio emission is
not related to a particular cluster member, but rather to the presence
of relativistic electrons (Lorentz factor $\gamma \gg$ 1000) and weak
magnetic fields ($\mu$G) in the intracluster space.

In this review, we focus on radio observations of this non-thermal
component in galaxy clusters. We outline our current knowledge on the
presence of non-thermal processes in the intracluster gas, and their
physical connections with the thermodynamical evolution of large-scale
structure. The relevance of the study of extended cluster radio
emission for cosmology is pointed out. On smaller scales, there are
only few indications of the possible presence of extended radio
emission in galaxy groups. These systems host diffuse $\sim$ 1 keV gas
called intragroup medium (IGM) \citep[see, e.g.,][]{Mulchaey96}. Radio
and hard X-ray emission possibly related to a non-thermal component of
the IGM has been recently pointed out by \citet{Delain06} and
\citet{Nakazawa07} respectively. The existence of diffuse radio
sources in galaxy groups has indeed to be tested with observations of
higher sensitivity.  Radio observations of the emission from
individual radio galaxies are not treated here. For a discussion on
cluster radio galaxies see the reviews by \citet{Feretti02a} and
\citet{Feretti07}. X-ray observations and simulations of the
non-thermal component in clusters are reviewed by \citet{Rephaeli08} -
Chapter 5, and \citet{Dolag08} - Chapter 15, this volume. The adopted
cosmology is $\Lambda$CDM (${\rm H}_0$=70 km ${\rm s}^{-1} {\rm
  Mpc}^{-1}$, $\Omega_{\rm m} = 0.3$, $\Omega_{\Lambda} = 0.7$).

\section{Extended radio emission in galaxy clusters}
\label{DiffExt}

The first detection of diffuse and extended radio emission in galaxy
clusters dates back to 1959, when \citeauthor{Large59} mapped for the
first time the Coma cluster at radio wavelengths, detecting an
extended radio source (Coma C) at its centre. The existence of Coma C
was later confirmed by \citet{Willson70}, who compared the single dish
data of \citet{Large59} with interferometric observations, and
determined that the observed radio emission was diffuse and not
associated with any cluster galaxy. From then on, high sensitivity
radio observations have revealed in about 50 clusters the existence of
diffuse non-thermal radio sources, not associated with active
galaxies but with the ICM. The power-law radio
spectrum\footnote{S$(\nu) \propto \nu^{-\alpha}$, see
  Eq. \ref{eq:sync}.} of this class of cluster sources indicates their
synchrotron nature, and thus the presence of relativistic electrons
(Lorentz factor $\gamma \gg$ 1000) and magnetic fields ($\sim0.1-1
\mu$G) permeating the cluster volume.

While the thermal gas emitting in X-rays is present in all clusters,
the detection of extended radio emission only in $\lesssim$ 10~\% of
the systems indicates that the non-thermal plasma is not a common
property of galaxy clusters. The very low surface brightness of
diffuse cluster radio sources makes them difficult to detect with
current radio telescopes. However, as we discuss here and in the
following sections, our current knowledge suggests that the lack of
radio emission in a high fraction of known clusters is not only
related to a limited sensitivity of the current
instruments\footnote{For instance, the 3$\sigma$ sensitivity limit for
  the NRAO VLA Sky Survey (NVSS) at 1.4 GHz, with a resolution of 45
  arcsec, is $\sim$1.35 mJy/beam \citep{Condon98}. The Very Large
  Array (VLA) is operated by the National Radio Astronomy
  Observatory.}, but also to physical reasons, as non-thermal
components over $\sim$ 1 Mpc scales are present only in the most
massive merging clusters. Discriminating between these two effects is
at present extremely difficult, and it will be one of the main goals
of future radio observations (see Sect. \ref{Discussion}).

The steep radio spectral index usually observed in diffuse cluster
radio sources ($\alpha \gtrsim$ 1) is indicative of ageing of the
emitting particles. The steepening of the electron spectrum is a
direct result of their Compton-synchrotron losses.  The highest energy
particles lose their energy more quickly. As a result, if cosmic rays
are produced in a single event with a power law energy distribution
\be
N(E) {\rm d}E = N_0 E^{-\delta} {\rm d}E \label{eq:ensp}
\ee
following the emergence of electrons from their sources (or
acceleration sites), their spectrum steepens as result of the shorter
energy loss time of high energy electrons. As a consequence the
synchrotron spectrum falls off rapidly beyond a certain break
frequency $\nu^*$, which shifts gradually to lower frequencies. If
instead particles are re-accelerated and/or continuously injected, as
suggested to occur in cluster diffuse sources, the energy spectrum of
relativistic particles may adopt more complex spectral shapes.

Apart from their common properties (nature of the emission, steep
radio spectra), diffuse and extended radio sources in clusters differ
in their physical properties, in particular: size, position in the
host cluster, intensity of polarised signal, morphology and
association to other cluster physical properties (e.g. dynamical
state, presence of a cooling flow). A working definition, that is
usually adopted, assigns cluster diffuse radio sources to three main
classes: halos, relics and mini-halos. Very schematically:

\begin{itemize}

\item {\sl radio halos} are extended ($\gtrsim$ 1 Mpc) diffuse radio
  sources at the centre of clusters, with a quite regular morphology,
  similar to the X-ray morphology of the system;

\item {\sl radio relics} have similar extensions and are also detected
  in merging clusters, but usually they are located in the cluster
  outskirts and they have an elongated morphology;

\item {\sl radio mini-halos} are smaller sources ($\lesssim$ 500 kpc)
  located at the centre of cooling flow clusters. They surround a
  powerful radio galaxy.

\end{itemize}

\noindent The main observational properties of these sources, useful
to test their different formation scenarios (Sect. \ref{Origin}), are
reviewed in more detail in the following (Sects. \ref{Halos},
\ref{MiniHalos} and \ref{Relics}).

\subsection{Radio halos}
\label{Halos}

\begin{figure*} 
\centering   
\includegraphics[width=0.7\textwidth]{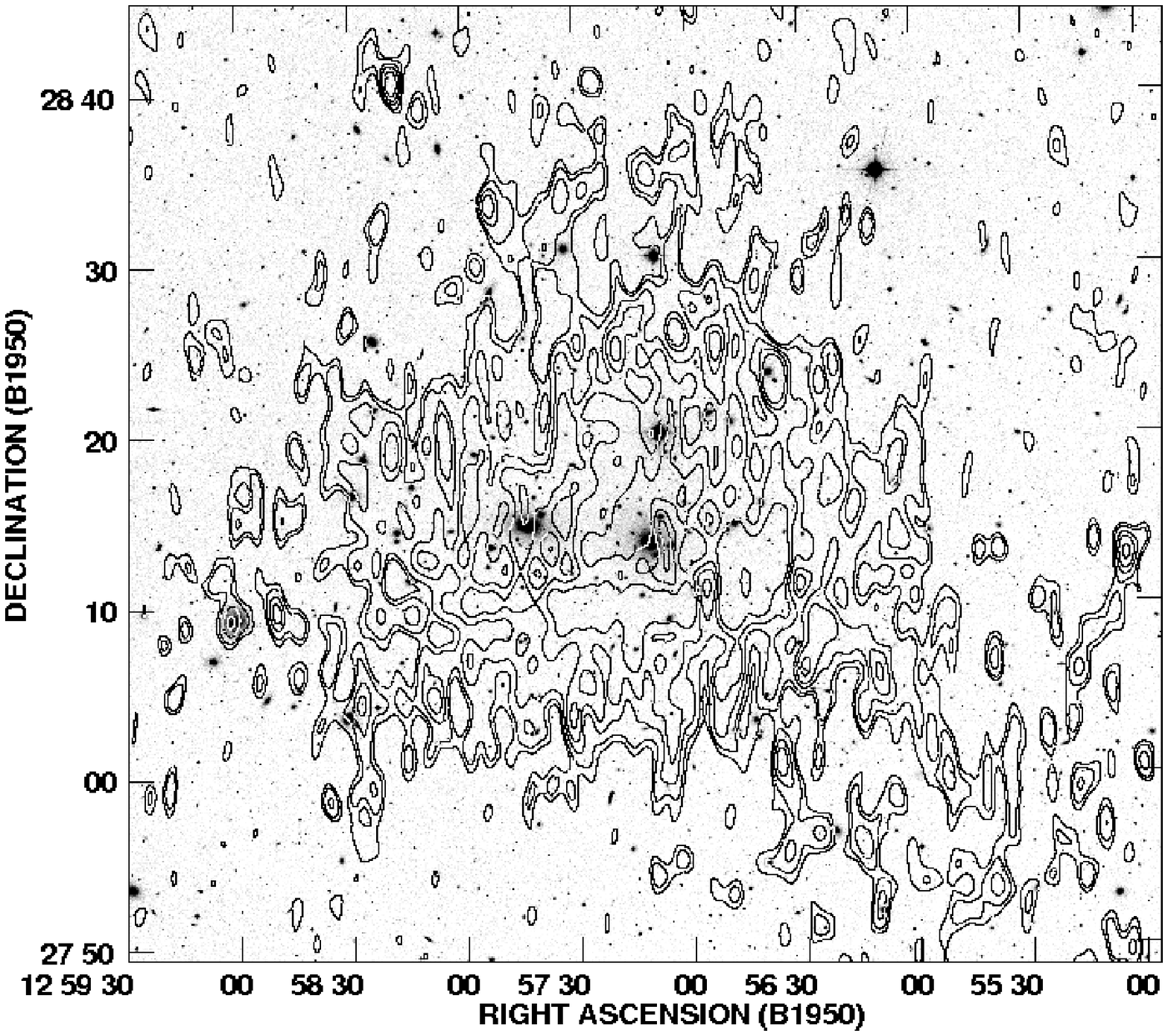}
\includegraphics[width=0.7\textwidth]{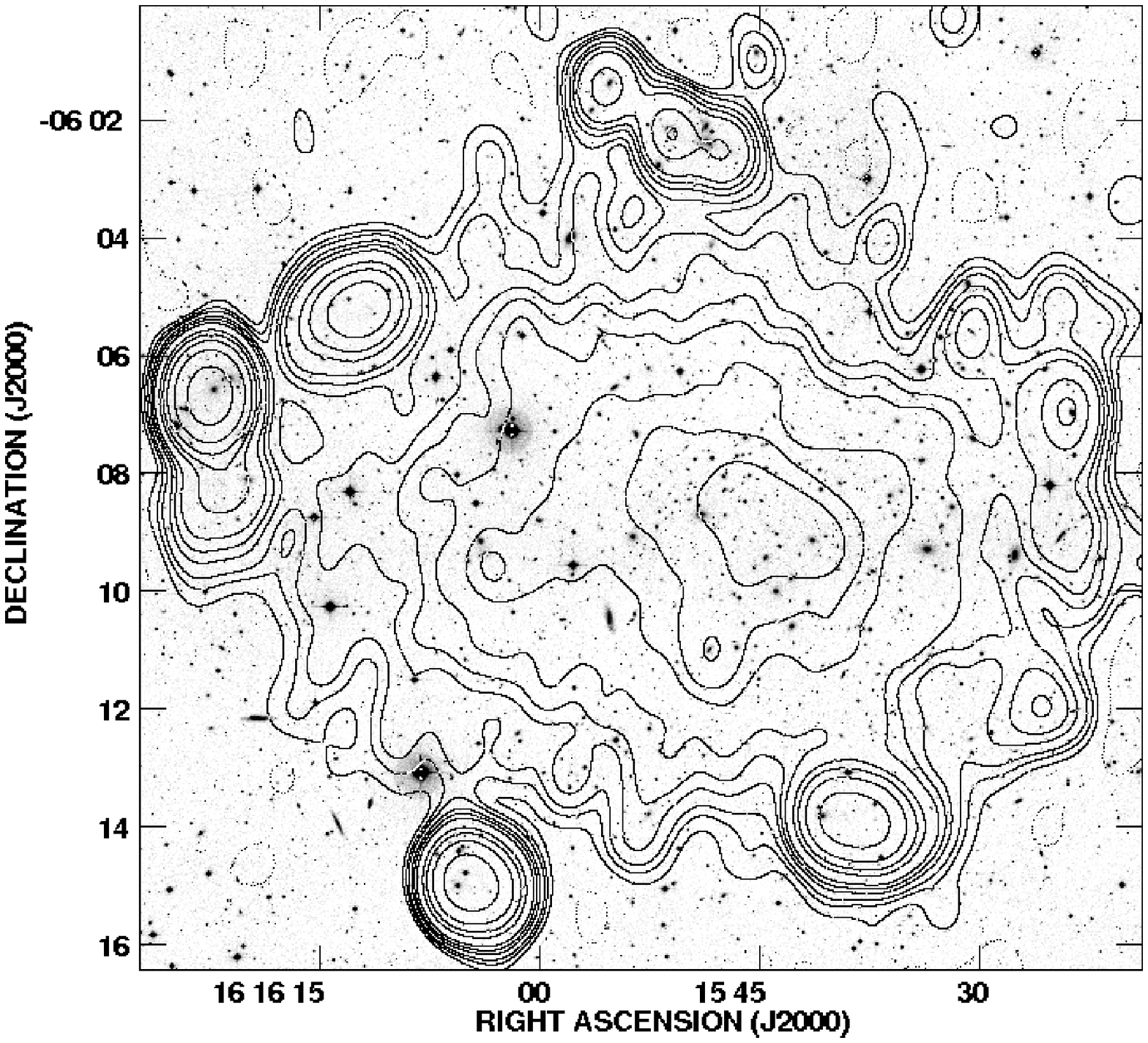}
\caption{{\bf Top:} 90 cm contours of the radio halo in the Coma
  cluster ($z$ = 0.023) are overlaid on the DSS optical image. Radio
  point sources have been subtracted \citep{Feretti02b}.  {\bf
    Bottom:} 20 cm radio contours \citep{Feretti01} overlaid on the
  deep, R-band image \citep{Maurogordato07} of the galaxy cluster
  A~2163 ($z$ = 0.203), hosting one of the most extended and powerful
  halos known so far.}
\label{fig:halos}
\end{figure*}

Coma C is the prototype of the low surface brightness ($\sim \mu{\rm
  Jy}$ arcsec$^{-2}$ at 1.4 GHz) and extended ($\gtrsim$ 1 Mpc) radio
sources permeating the central volume of clusters, usually referred as
``radio halos''. Their radio morphology is quite regular (see
Fig. \ref{fig:halos}) and their radio emission is unpolarised down to
a few percent level. The first (and only) successful detection of
polarised emission from a radio halo has been published recently by
\citet{Govoni05}.  One can find a compilation of most of the currently
known radio halos in the recent review by \citet{Feretti07}.

\begin{figure}    
\centering
\includegraphics[width=0.7\textwidth]{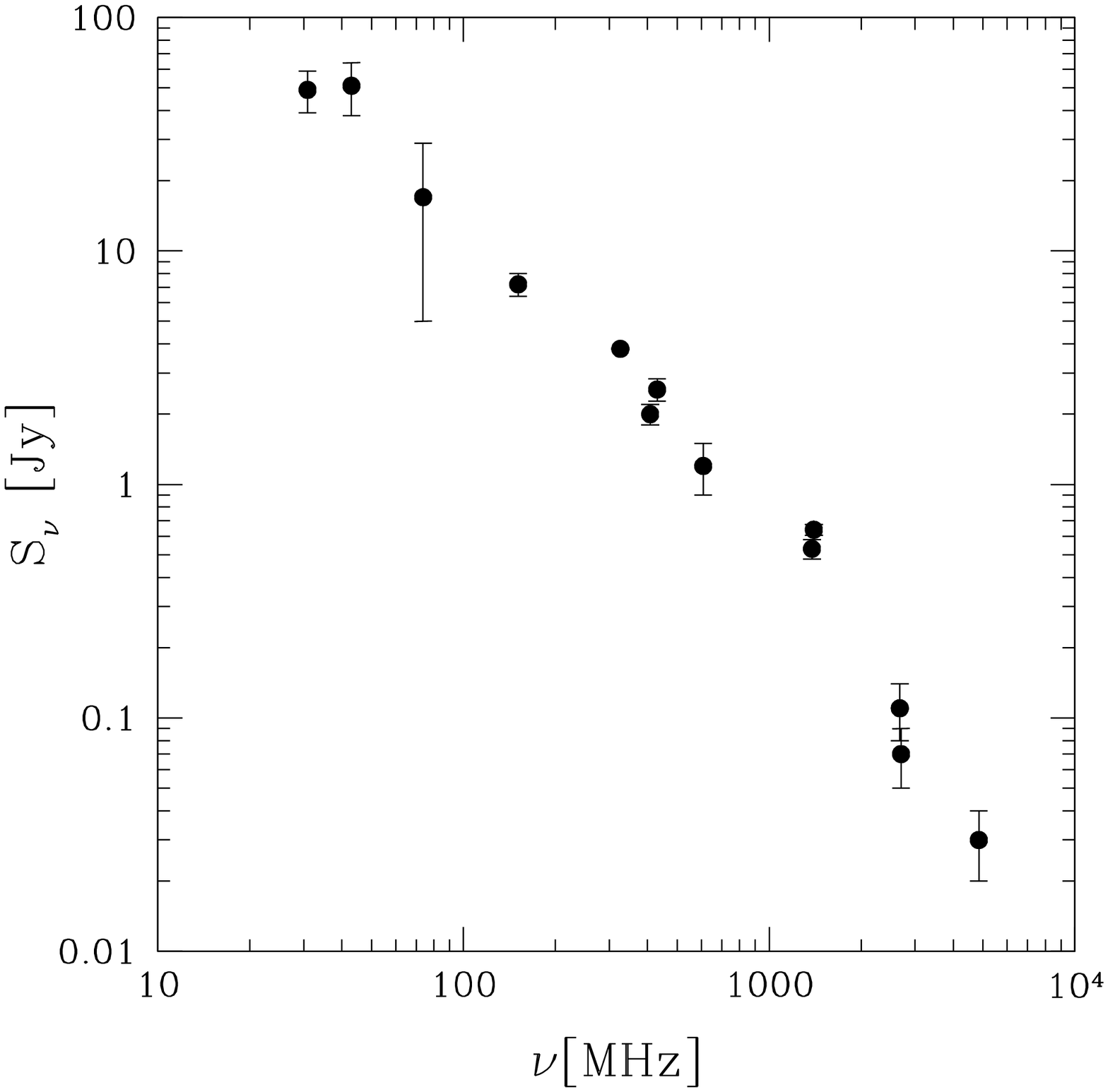}
\caption{Spectrum of the radio halo in the Coma cluster (Coma C). A
  steepening in the spectrum stands out clearly at $\nu >$ 1 GHz
  \citep[adapted from][]{Thierbach03}.}
\label{fig:specComa}
\end{figure}

Spectral index studies of extended radio sources can give important
hints on the energy spectrum of relativistic electrons and, due to the
dependence of the synchrotron emissivity on the magnetic field
intensity (Eq. \ref{eq:sync}), on the magnetic field distribution
\citep{Brunetti01}. In recent years, many observational efforts have
thus been devoted to multi-frequency observations of radio halos, in
order to get more and more accurate determinations of the integrated
radio spectrum and, possibly, of spatially resolved spectral index
maps. These studies are limited however by the capability of current
instruments to do multi-frequency observations at the sensitivity
required for studying radio halos ($\sim {\rm mJy} - \mu{\rm Jy}$
arcsec$^{-2}$ going from the MHz to the GHz range). In a few cases, a
steepening of the halo spectrum at high frequency (as in
Fig. \ref{fig:specComa} in the case of the Coma cluster) has been
detected (A~2319: \citealp{Feretti97}; Coma: \citealp{Thierbach03};
A~754: \citealp{Bacchi03}; A~3562: \citealp{Giacintucci05}). Indications
that the spectral index steepens radially with the distance from the
cluster centre have been pointed out by \citet{Giovannini93} in Coma,
and \citet{Feretti04a} in A~665 and A~2163. More recently,
\citet{Orru07} have shown that the radio halos in A~2744 and A~2219 have
a mean spectral index, averaged over the whole cluster, without a
clear radial steepening, but with a very patchy structure. Very
interestingly, their radio/X-ray comparison shows flatter spectral
indexes in regions characterised by higher ICM temperature.

Current observational results suggest other strong connections between
the physical properties of radio halos and of their host clusters.
All radio halos discovered up to now are at the centre of clusters
with signatures of a disturbed dynamical state and without a cooling
core. However, not all merging clusters host a radio halo. The
detection rate of radio halos is actually quite low: 5~\% in a complete
cluster sample at the detection limit of the NVSS, which grows to
$\sim$ 35~\% when only the most luminous X-ray clusters are considered
($L_{X[0.1-2.4~{\rm keV}]} > 0.6 \times 10^{45}~{h_{70}}^{-2}~{\rm
  erg~{\rm s}^{-1}}$) \citep{Giovannini99, Giovannini02}.
The fact that radio and X-ray properties of clusters are connected is
also suggested by a close similarity of the morphology of radio halos
and the X-ray emission of their host clusters. This has firstly been
revealed in a qualitative way \citep{Deiss97, Feretti99a, Liang00},
and afterwards quantitatively confirmed by the relation between the
point-to-point surface brightness of the cluster radio and X-ray
emission \citep{Govoni01c, Feretti01}.

\begin{figure*}   
\centering
\includegraphics[width=0.45\textwidth]{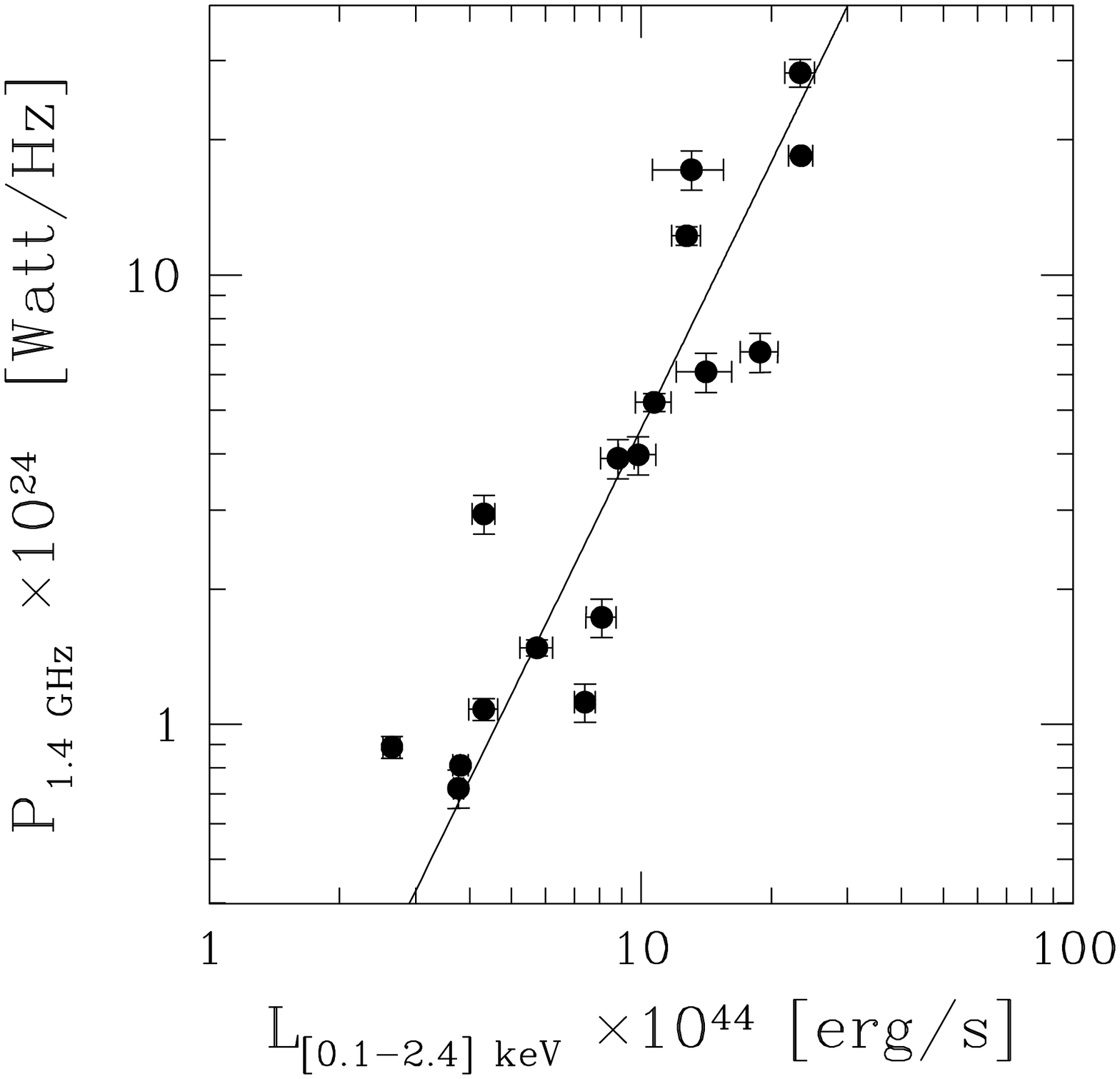}
\includegraphics[width=0.45\textwidth]{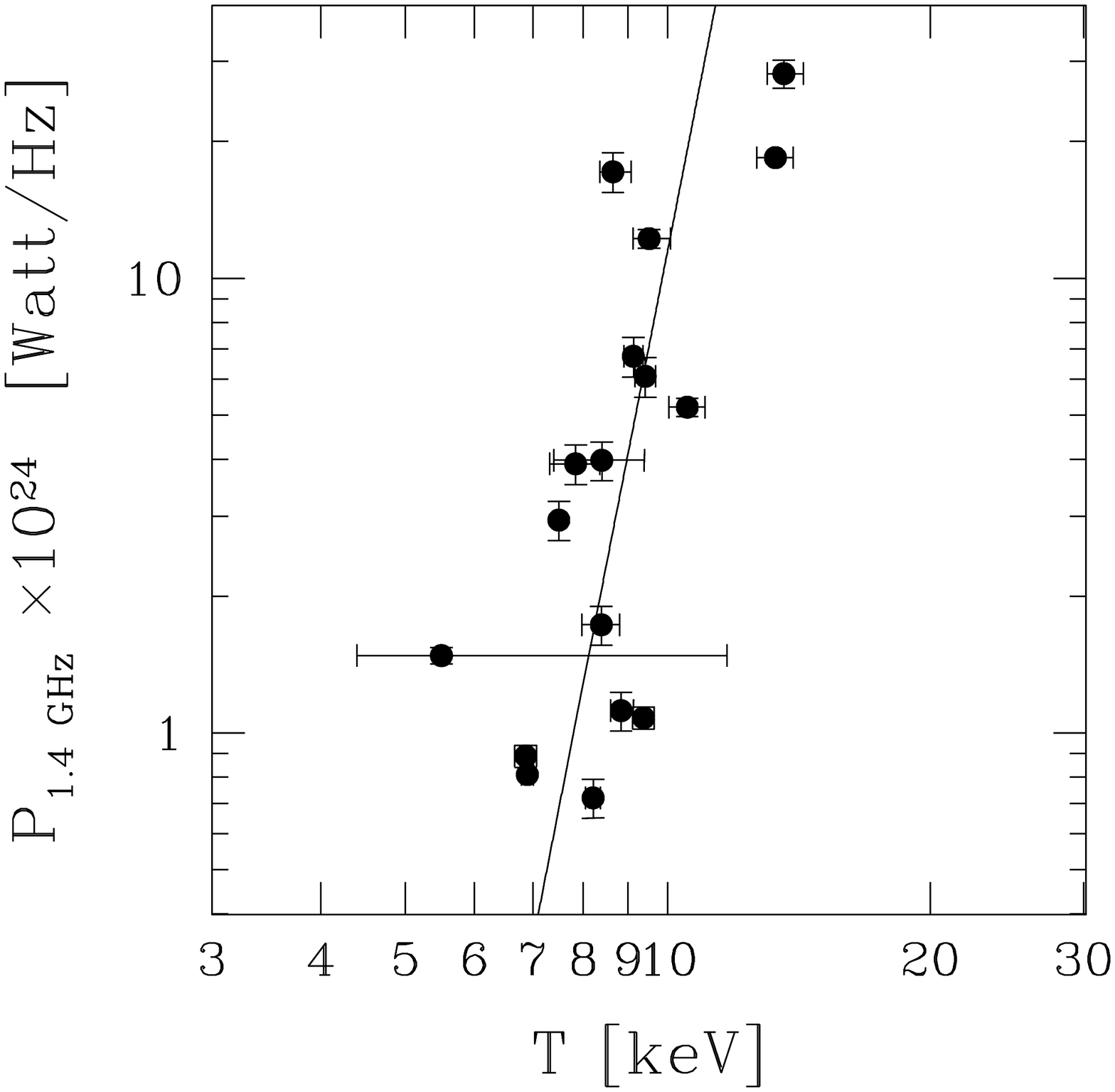}
\caption{Radio power at 1.4 GHz of giant ($\gtrsim$ 1 Mpc) radio halos
  vs. a) {\bf left:} cluster X-ray luminosity, and b) {\bf right:}
  cluster X-ray temperature \citep[adapted from][]{Cassano06}.}
\label{fig:LXLRTX}
\end{figure*}

Additionally, a strong correlation has been pointed out between the
radio power ($P_{\nu}$) of halos and the X-ray luminosity ($L_X$) of
their host clusters (e.g. \citealt{Liang00, Giovannini02, Ensslin02,
  Cassano06}; see left panel of Fig. \ref{fig:LXLRTX}). A relation
with a much larger scatter between radio power and X-ray temperature
of the ICM ($T_X$) has also been suggested
\citep[e.g.][]{Colafrancesco99, Liang00} (see right panel of
Fig. \ref{fig:LXLRTX}). Since both the X-ray luminosity and
temperature of clusters correlate with mass \citep[e.g.][]{Neumann99,
  Neumann01}, the observed $P_{\nu}$ - $L_X$ and $P_{\nu}$ - $T_X$
relation could reflect a dependence of the radio halo luminosity on
the cluster mass, with interesting implications on the theoretical
models of cosmic ray production (see Sect. \ref{Origin}). Current
results suggest $P_{\rm 1.4~GHz } \propto M^{a}$, with $a$ = 2.3 or
larger, depending on the methods applied to estimate the cluster mass
(see \citealp{Feretti07} and references therein).

Recently, \citet{Cassano07} pointed out that the fraction of the radio
emitting cluster volume significantly increases with the cluster
mass. This break of self-similarity can give important constraints on
the physical parameters entering the hierarchical formation scenario,
since it suggests that the distributions of the magnetic field and
relativistic electrons change with the cluster mass.

\subsection{Radio mini-halos}
\label{MiniHalos}

\begin{figure*}    
\centering
\includegraphics[width=1\textwidth]{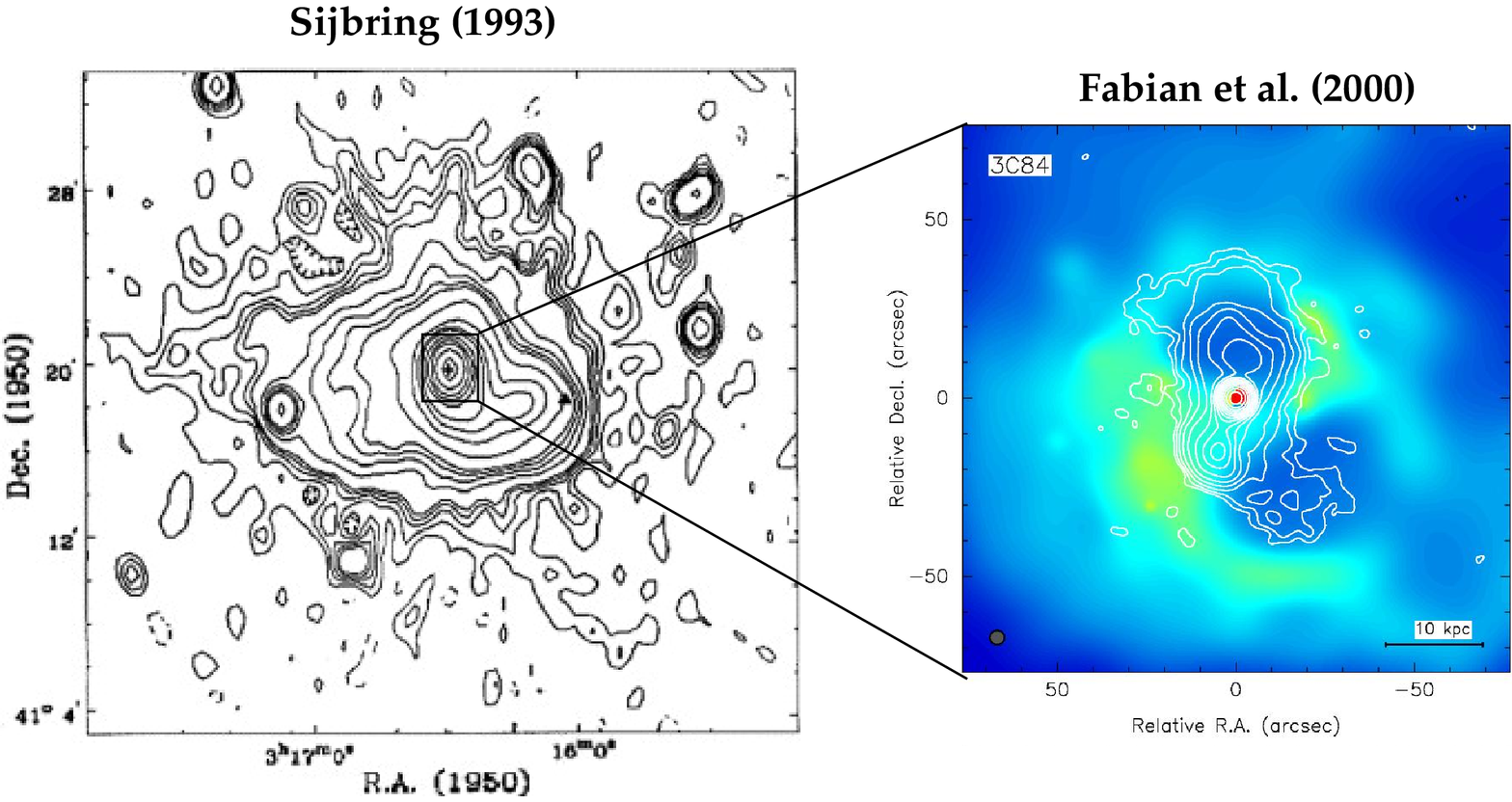}
\includegraphics[width=0.5\textwidth]{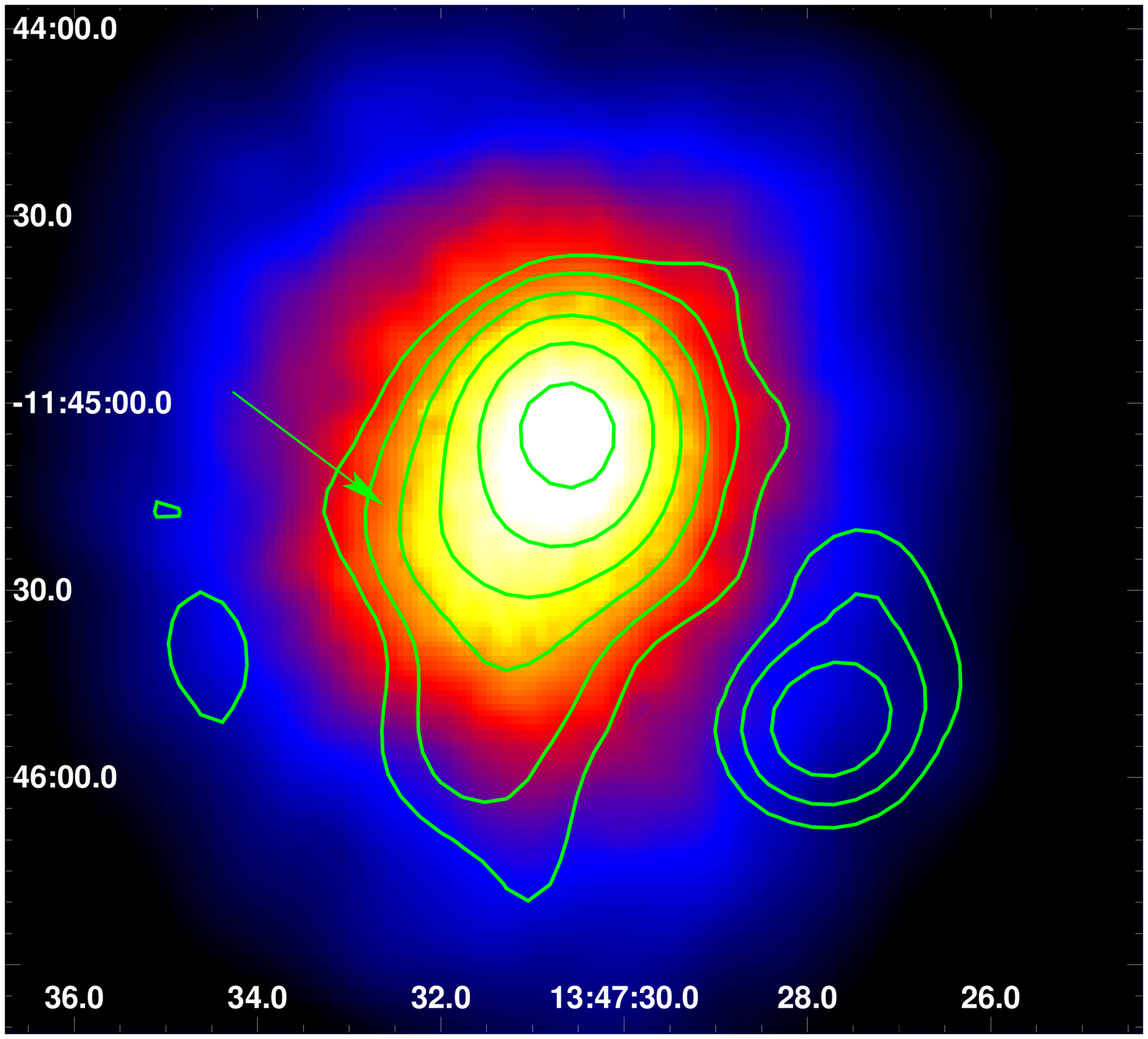}
\caption{{\bf Top:} 327 MHz map of the mini-halo in the Perseus
  Cluster ($z$ = 0.018). The source is centred on the position of the
  cD galaxy NGC 1275 (indicated with a cross). The inset shows radio
  contours overlaid on the X-ray image of the central $\sim 1^{\prime}$ region
  of Perseus. The holes evident in the X-ray emission are due to
  subsonic expansion of the buoyant radio lobes of the central radio
  galaxy 3C~84 (adapted from \citealt{Sijbring93} and \citealt{Fabian00}).
 {\bf Bottom:} 1.4 GHz contours of the radio mini-halo (the most distant
  ever detected) in the galaxy cluster RXJ 1347.5$-$1145 ($z$ = 0.451),
  superimposed on the {\sl XMM-Newton} image of the cluster. The green
  arrow indicates an elongation in the radio mini-halo morphology,
  corresponding to a sub-structure detected in X-rays \citep[adapted
    from][]{Gitti07a}.}
\label{fig:minihalos}
\end{figure*}

The so-called ``radio mini-halos'' (Fig. \ref{fig:minihalos}) differ
from the above described radio halos not only because of their smaller
size (few $10^2$ kpc), as their name suggests, but also in the typical
properties of their host clusters. Actually, mini-halos are diffuse
radio sources with a steep spectral index, which are found around
powerful radio galaxies at the centre of cooling core clusters. The
total size of mini-halos is comparable to that of the cooling
region. Since major mergers are able to disrupt (or at least disturb)
cluster cooling flows \citep[e.g.][]{Buote96, Gomez02}, the main
physical difference between giant and mini-halos is that they are
hosted in clusters with and with no evidence of {\sl major} mergers
respectively.

However, recent results revealed the existence of two cooling flow
clusters with signatures of merging activity in the central region and
hosting a radio mini-halo: A~2142 \citep{Giovannini00} and RXJ
1347.5$-$1145 \citep[][see bottom panel of
  Fig. \ref{fig:minihalos}]{Gitti07a}. Contrary to what expected in
relaxed systems, both the clusters are dominated by two brightest
cluster galaxies (BCGs). In A~2142, the central cooling flow has been
disturbed but not destroyed by an unequal merger, observed $1-2$~Gyr
after the initial core crossing \citep{Markevitch00}. The cooling flow
in RXJ 1347.5$-$1145 is one of the most massive ever detected,
suggesting a relatively long interval of time in which the central
part of the cluster has evolved undisturbed to a nearly relaxed state
\citep{Gitti07b}. The X-ray analysis of \citet{Gitti07b} reveals
however a sub-structure in the south-east part of the cluster,
corresponding to an elongation in the radio mini-halo morphology (see
bottom panel of Fig.~\ref{fig:minihalos}). Indications of possible {\it
  minor} mergers have been detected also in other clusters hosting
radio mini halos (Perseus: \citealp{Ettori98, Furusho01}; A~2390:
\citealp{Allen01}; A~2626: \citealp{Mohr96}). In these cases, however,
the merging substructures are located well outside the diffuse radio
source.

Several radio halos have been discovered in radio surveys
\citep[e.g. NVSS,][]{Giovannini99}, where their relatively large beam
provides the necessary signal-to-noise ratio to spot these elusive
sources. Due to their extremely low surface brightness and large
angular extent radio halos are actually best studied at low spatial
resolution.  Unfortunately, due to the combination of their smaller
angular size and the strong radio emission of the central radio
galaxy, the detection of a mini-halo requires a much higher dynamic
range and resolution than those in available surveys, and this
complicates their detection. As a consequence, our current
observational knowledge on mini-halos is limited to less than ten
known sources (PKS 0745$-$191: \citealp{Baum91}; Perseus:
\citealp{Burns92}; A~2142: \citealp{Giovannini00}; A2390:
\citealp{Bacchi03}; A~2626: \citealp{Rizza00, Gitti04}; RXJ
1347.5$-$1145: \citealp{Gitti07a}). This, together with the peculiar
properties of A~2142 and RXJ 1347.5$-$1145, and the possible connection
between radio mini-halos and minor cluster mergers, requires better
statistics to test the current theoretical models on the origin of
their radio emission, which are discussed in Sect. \ref{Origin}.

\subsection{Radio relics}
\label{Relics}

As clearly stated by \citet{Kempner04a} there is quite a lot of
confusion in the literature when speaking about ``radio relics''. This
definition is actually adopted for at least three different kinds of
radio sources in galaxy clusters, characterised by significantly
different observational properties. Certain features are in common for
all the different kinds of relic sources, such as their steep radio
spectrum ($\alpha \gtrsim$ 1) and a general filamentary morphology.

\begin{figure*}    
\centering
\includegraphics[width=0.6\textwidth]{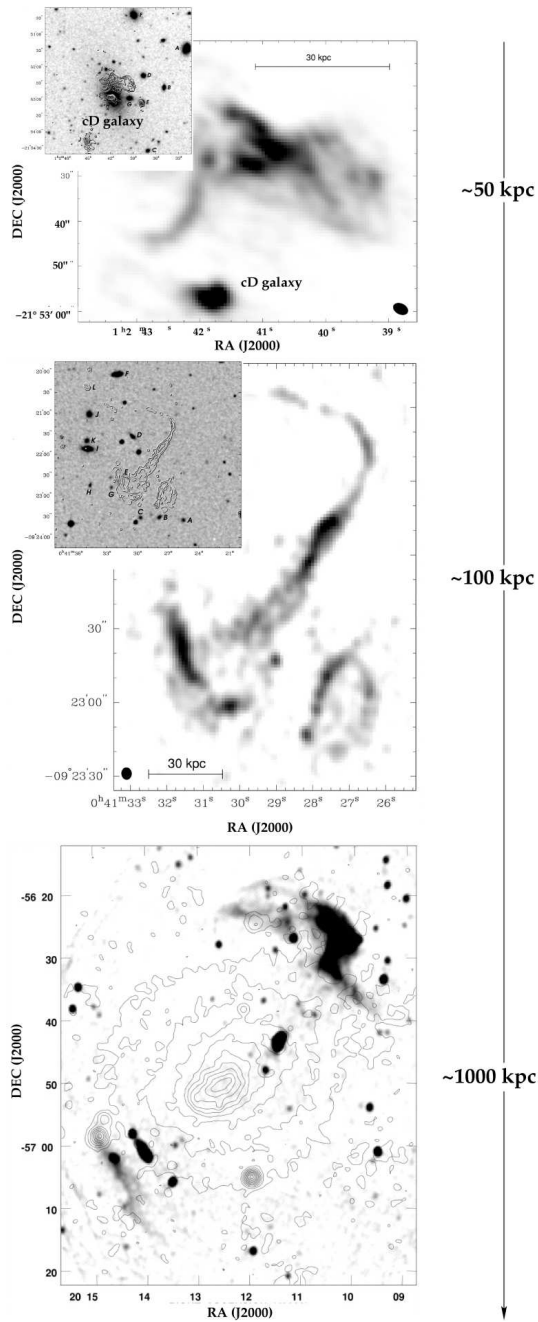}
\caption{Examples of different diffuse radio sources in clusters
  classified as ``relics'' in the literature. From top to bottom, an
  ``AGN relic'', a ``Phoenix'' and a ``Radio Gischt'' (see
  \citealp{Kempner04a} and Sects. \ref{Relics} and \ref{Origin}). {\bf
    Top panel:} VLA image at 1.4 GHz of the ``AGN relic'' source in
  A~133 ($z$ = 0.056), close to the radio emitting cD galaxy at the
  cluster centroid. In the inset, radio contours are superimposed on
  the optical, DSS-2 image of the area \citep[adapted
    from][]{Slee01}. {\bf Middle panel:} VLA image at 1.4 GHz of the
  ``Phoenix'' source in the periphery of A~85 ($z$ = 0.055) \citep[see
    also Fig. 11 in][]{Slee01}. As before, the inset shows radio
  contours superimposed on the optical DSS-2 image of the region
  around the radio source \citep[adapted from][]{Slee01}. {\bf Bottom
    panel:} X-ray contours (ROSAT data, $0.1-2.4$ keV energy band)
  overlaid on the 843 MHz Molonglo Observatory Synthesis Telescope
  (MOST) image of A~3667 ($z$ = 0.056) \citep[from][]{Roettgering97}.}
\label{fig:relics}
\end{figure*}

A first group of sources (see an example in the top panel in
Fig. \ref{fig:relics}) has typical sizes of several 10 kpc and
low/intermediate polarisation intensity ($\lesssim$ 20~\%). They are
generally located in the central cluster regions, close to the cD
galaxy, often showing an anti-correlation with the ICM
density. Actually, in some clusters, relic as well as AGN radio
emission has been detected inside holes in the central X-ray emission
of the thermal gas (see for instance the inset in the top panel of
Fig. \ref{fig:minihalos}). These cavities are actually related to the
cyclic outburst activity of the central AGN. The cases in which no
radio emission has been detected, or has been revealed only by low
frequency radio observations \citep[e.g. the Perseus outer cavities
  discovered by][]{Fabian02}, are due to buoyant old radio lobes,
whose spectrum is too steep to be detected in the GHz range. A more
detailed discussion on these ``radio ghosts'' and X-ray cavities in
clusters can be found in a recent review by \citet{McNamara07}.

Both the second and the third class of ``cluster relics'' (middle and
bottom panel of Fig. \ref{fig:relics}) are strongly correlated with
the ICM properties. They are commonly found in merging clusters, and,
in some cases, a spatial correlation with shocks in the thermal gas
has been pointed out \citep[e.g.][]{Kassim01}. Both of these classes
of sources do not have a likely parent radio galaxy nearby, are
generally polarised at the level of about $10-30$~\% and located in the
cluster periphery. While the (rare) objects in the second group are
characterised by intermediate linear scales compared to the other two
(with typical sizes of $\sim 10^2$ kpc, e.g. \citealp{Slee01}), the
most extended radio emission among the ``relic sources'' comes from
the third class.

These giant relics, with extensions ranging from a few $\sim 10^2$ kpc
to $10^3$ kpc, have been detected in merging clusters both with and
without cooling cores. The major axis of their elongated structure is
generally nearly perpendicular to the direction of the cluster
radius. Some of the most extended and powerful giant relics are
located in clusters with central radio halos (e.g. A~2256:
\citealp{Clarke06}).  In a few cases, two symmetric relics have been
observed, as in the bottom panel of Fig. \ref{fig:relics} (A~3667:
\citealp{Roettgering97, Johnston02}; A~3376: \citealp{Bagchi06}; A~1240:
\citealp{Kempner01}). More ``exotic'' giant radio relics have also
been discovered, such as sources located in poor clusters (Abell S~753:
\citealp{Subrahmanyan03}), far away from the cluster centre (A~786:
\citealp{Giovannini00}), or even in intracluster filaments of galaxies
(ZwCl~2341.1+0000: \citealp{Bagchi02}). Relics elongated from the
cluster centre to the periphery (A~115: \citealp{Govoni01b}), or with a
circular shape (A~1664: \citealp{Feretti05}) have also been
detected. Similarly to radio halos (Sect \ref{Halos}), the detection
rate of giant radio relics at the sensitivity limit of NVSS is $\sim$
6~\% \citep{Giovannini02}, and the relic radio power correlates with
the cluster X-ray luminosity \citep{Feretti02b}, even though with a
larger scatter.

\section{Origin of radio emitting particles}
\label{Origin}

Based on the observational results summarised in Table
\ref{tab:summary1} and in the previous sections, and on the
theoretical models reviewed, for instance, by \citet{Brunetti04a},
\citet{Blasi07} and \citet{Dolag08}, we have now a formation scenario
for the different diffuse and extended radio sources in clusters. The
current theories on the origin of the non-thermal component in galaxy
clusters will be the starting point for new observational studies with
the next generation radio telescopes (Sect. \ref{Discussion}).

\begin{table*}
\centering
\caption{Main observational properties of the different sources of
  diffuse radio emission in galaxy clusters. Note that linear
  polarisation levels of $\sim 20-40$~\% have been detected in
  filamentary structures of the radio halo in A~2255 by
  \citet{Govoni05}.}
\label{tab:summary1}       
\begin{tabular}{llllll}
\hline\noalign{\smallskip}
Type & Position & Size & $\alpha$ & Polarisation & Example  \\
\noalign{\smallskip}\hline\noalign{\smallskip}
Halo & Centrally & $\gtrsim$ Mpc & $\gtrsim$ 1 & $<$ few \% & Coma \\
 & peaked & & & & \\
Giant relic & Peripheral & $\sim$ Mpc & $\gtrsim$ 1 & $\sim 10-30$~\% & Abell 3667 \\
 & & & & & \\
Mini-halo & Centrally & $\lesssim$ 0.5 Mpc & $\gtrsim$ 1.5 &  $<$ few \% & Perseus \\
 & peaked & & & & \\
Phoenix & Peripheral & $\sim 10^2$ kpc & $\gtrsim$ 1.5 &  $\sim 10-30$~\% & Abell 85 \\
 & & & & & \\
AGN relic & Close to the & few $\times$ 10 kpc & $\gtrsim$ 1.5 & $\lesssim$ 20~\% & Abell 133 \\
 & host galaxy & & & & \\ 
\noalign{\smallskip}\hline
\end{tabular}
\end{table*}

Giant radio halos and relics are the most spectacular radio sources in
clusters, and, as stated above, their synchrotron spectrum indicates
the presence of cosmic rays that gyrate around magnetic field lines,
frozen in the ICM. Therefore, relativistic particles cannot stream out
from the gravitational field of the cluster, but they can still
diffuse along magnetic field lines.  It has been shown however
\citep{Voelk96, Berezinsky97, Voelk99} that typical relativistic
electrons in radio halos and relics (with $\gamma \sim 1000 - 5000$)
have diffusion times which are longer than the Hubble time. They could
therefore be simply diffused over cluster scales from one or more
active radio galaxies \citep{Jaffe77, Rephaeli77}.  However, the steep
radio spectra of these sources indicate short lifetimes for the
radiating particles ($\sim 10^8$ yr), which lose energy not only via
synchrotron emission, but also due to interactions with the Cosmic
Microwave Background (CMB) photons (via Compton scattering emission)
and with the ICM (via Coulomb interactions and Bremsstrahlung
emission). The main radiative losses of electrons are due to Compton
scattering of the CMB, and synchrotron emission; the former process
dominates for $B < 3 \mu$G (the field equivalent of the CMB energy
density). The radiative lifetime of a relativistic lepton with a
Lorentz factor $\gamma < 10^8$ is thus approximately given by
\citep[e.g.][]{Longair81, Meisenheimer89}
\be 
\tau
  \approx 2 \times 10^{12} \gamma^{-1} \left[(1+z)^4 + \left
    (\frac{B}{3.3 {\mu}{\rm G}}\right)^2 \right]^{-1} {\rm years}.
\ee
Since the expected diffusion velocity of the relativistic
electrons is of the order of 100 km~${\rm s}^{-1}$ (Alfv\'en speed),
cosmic rays do not have the time to propagate over the Mpc-scales of
giant cluster radio sources. This excludes the hypothesis that
relativistic electrons are produced at a localised point, and requires
{\sl in situ} acceleration mechanisms. Basically, two classes of
models have been proposed:

\begin{itemize}

\item the {\sl primary models}, which predict that electrons are
  accelerated by shocks and/or turbulence induced during cluster
  mergers. Shocks can (re-)accelerate particles via Fermi-I processes
  \citep{Ensslin98} or adiabatic compression of fossil radio plasma
  \citep{Ensslin01}; turbulence via stochastic, Fermi-II processes
  \citep{Brunetti01} or magnetohydrodynamic (MHD) waves (Alfv\'en
  waves: \citealp{Brunetti04b}; magneto-sonic waves:
  \citealp{Cassano05});

\item the {\sl secondary models}, in which relativistic electrons are
  continuously injected by hadronic collisions between the thermal
  ions of the ICM and relativistic protons, the latter (characterised
  by significantly longer lifetime compared to relativistic electrons)
  having been accelerated during the whole cluster history
  \citep[e.g.][]{Dennison80, Blasi99, Dolag00a}.

\end{itemize}

\noindent The observational properties of radio halos and relics (see
Sects. \ref{Halos} and \ref{Relics}) are more in favour of primary
models. The strongest point leading to this conclusion is the fact
that diffuse and extended radio emission has been detected up to now
only in merging clusters. A strong connection between cluster mergers
and cosmic ray production is required in primary models, and is not
expected in secondary models. In this respect, the fact that halos and
relics are quite rare in clusters is again disfavouring the hadronic
collision hypothesis, based on which we should expect electron
acceleration to be possible in all galaxy clusters.

Since the shape of the synchrotron spectrum depends on the last
acceleration phase of cosmic rays, detailed studies of the spectral
index distribution in radio halos and relics provide important
information on acceleration mechanisms acting in clusters. Primary
models predict a radial steepening and a complex spatial distribution
of the spectral index $\alpha$, due to the existence of a) a maximum
energy to which electrons can be accelerated 
\citep[$\gamma <  10^5$,][and references therein]{Blasi04}, 
and b) different
re-acceleration processes in different cluster regions. Secondary
models assume that cosmic ray protons are accelerated during structure
formation over cosmological epochs and accumulated in clusters. The
collision of these protons with the thermal ICM would continuously
inject electrons, resulting in a spectral index distribution unrelated
to the intracluster magnetic field strength, and thus not dependent on
the position in the cluster. The radial spectral steepening and/or the
patchy structure of spectral index maps observed in several radio
halos (Sect. \ref{Halos}) are clearly favouring primary models.

The hadronic collision hypothesis predicts power-law spectra flatter
than primary models ($\alpha \lesssim 1.5$) and magnetic field values
higher than a few $\mu$G. Observational results are controversial
concerning these points, due to the observed intermediate values of
the radio spectral index ($\alpha \sim 1 - 1.5$), and the widely
differing estimates for mean intracluster magnetic field values
(Sect. \ref{Magnetic}).  Finally, emission of gamma-rays is expected
in secondary models \citep{Blasi07}, a challenging point to be tested
observationally (\citealp{Rephaeli08} - Chapter 5, this volume). On
the other hand, it has been suggested \citep{Bykov00, Miniati03} that
the detection of gamma-ray emission from clusters may not necessarily
reflect the hadronic origin of cosmic rays, since it could be related
to the Compton scattering of CMB photon from shock-accelerated,
intracluster electrons.

Given our current observational and theoretical knowledge, cosmic rays
in giant radio relics (bottom panel of Fig. \ref{fig:relics}) are most
likely originating from Fermi-I diffuse shock acceleration of ICM
electrons (e.g. \citealp{Hoeft07, Bykov08} - Chapter 7, this
volume). These radio sources would therefore
trace the rim of shock fronts resulting from cluster mergers in the
ICM, and they have been named ``radio gischt\footnote{German for the
  spray on the tops of ocean waves.}'' by \citet{Kempner04a}.
Firstly, this hypothesis is in agreement with the morphology and the
position of most of the detected giant relics, which appear as
elongated, sometimes symmetric, radio sources in the cluster
periphery, where we expect to find arc-like shock fronts resulting
from major cluster mergers \citep[e.g.][]{Schindler02}. Secondly, the
quite strong linear polarisation detected in giant relics would be in
agreement with the model prediction of magnetic fields aligned with
the shock front. Based on some observational results, however, a clear
association between shocks and giant radio relics is not always
straightforward. This is true in the case of the ``exotic'' giant
relics mentioned in Sect. \ref{Relics} (e.g. those with circular
shapes, or located in intracluster filaments). Additionally,
\citet{Feretti06} did not detect a shock wave corresponding to the
radio relic in the Coma cluster. They suggested that, similarly to
radio halos (see below), the radio emission of this relic source is
instead related to turbulence in the ICM. Currently, the main
observational limitation to test the origin of giant radio relics
comes from X-ray data. The sensitivity of X-ray instruments is not
high enough to detect shock waves in the external regions of clusters,
where the gas density and thus the X-ray surface brightness are very
low, and where most of radio relics have been detected (see for
instance the radio relic found in A~521 by \citet{Ferrari03a}; see also
\citealt{Ferrari06a, Giacintucci06}).

The second class of relic sources pointed out in Sect. \ref{Relics}
(middle panel in Fig. \ref{fig:relics}), characterised by smaller
sizes than giant relics ($\sim 10^2$ kpc vs. $10^3$ kpc), are most
likely originating from adiabatic compression in cluster shocks of
fossil radio plasma, released by an AGN whose central engine has
ceased to inject fresh plasma $\lesssim$ 1 Gyr ago \citep{Ensslin01}.
The old non-thermal electrons, that would be undetectable at high
($\sim$ GHz) frequencies, are actually re-energised by the shock. This
class of sources are therefore also called ``Phoenix''
\citep{Kempner04a}. The main physical difference with giant relics is
related to the fact that, in the latter, shocks accelerate thermal ICM
electrons to relativistic velocities through Fermi-I processes, while,
in the case of Phoenix sources, the shock waves energise the
relativistic plasma by adiabatic compression. The sound speed inside
the fossil radio plasma is actually so high that shock waves cannot
penetrate into the radio cocoons. These sources are rare because they
require shocks and fossil plasma in the same region of the
cluster. Moreover, adiabatic compression is efficient in
re-accelerating electrons only if the time elapsed since the last AGN
activity is not too long, i.e. less than about 0.1 Gyr in the cluster
centre and less than about 1 Gyr in the periphery. All this also
explains why we detect Phoenix sources in the external regions of
clusters.

Among the sources called ``radio relics'' in the literature, only the
smallest (several tens of kpc) are real ``AGN relics'' (top panel in
Fig. \ref{fig:relics}). They are extinct or dying AGNs, in which the
central nucleus has switched off, leaving the radio plasma to evolve
passively \citep[e.g.][]{Murgia05, Parma07}. Their spectrum becomes
steeper and steeper, making the source more and more difficult to be
detected at high frequencies, until it disappears completely (see
Sect. \ref{Relics}). Due to the short radiative lifetime of their
electrons ($\sim 10^7 - 10^8$ yrs), these sources are usually located
close to their host galaxy, which did not have time enough to move far
away in the cluster potential.

In the case of giant radio halos, spectral index maps show no evidence
of flattening at the location of shocks detected in X-rays (A~665:
\citealp{Feretti04a, Markevitch01}; A~2744: \citealp{Orru07,
  Kempner04b}).  This agrees with theoretical results showing that
shocks in major mergers are too weak to produce relativistic particles
uniformly over the whole central $\sim$ 1 Mpc area of clusters
\citep{Gabici03}. Although it cannot be excluded that shock
acceleration may be efficient in some particular regions of a halo
\citep[e.g.][]{Markevitch05}, it has been suggested that cluster
turbulence generated by cluster mergers may efficiently accelerate
electrons in the cluster volume \citep[e.g.][]{Cassano05}. The
observed steepening of the spectral index with the distance from the
cluster centre, and the few available spectral index maps showing
flatter spectra in the regions influenced by merger processes
(Sect. \ref{Halos}) support the scenario that ICM turbulence supplies
the energy for the radiating electrons.

However, if the predictions of primary models better agree with the
observational results, the merging event cannot be solely responsible
for electron re-acceleration in giant radio halos and relics, because
$\gtrsim$ 40~\% of clusters show evidence of a disturbed dynamical
state \citep{Jones99}, while only $\lesssim$ 10~\% possess radio halos
and/or relics. As we have seen in Sect. \ref{Halos}, the power of
observed radio halos $P_{\nu}$ seems to correlate with the mass $M$ of
their host cluster. The energy available to accelerate relativistic
particles during cluster mergers is a fraction of the gravitational
potential energy released during the merging event, that in turn
scales as $\sim M^2$. The $P_{\nu}$ - $M$ relation could thus suggest
that only the most massive mergers are energetic enough to efficiently
accelerate cosmic rays \citep{Buote01}. A recent model by
\citet{Cassano05} is in agreement with this conclusion, showing that
only massive clusters can host giant radio halos. The probability to
form these extended radio sources increases drastically for cluster
masses $M \geq 2 \times 10^{15}$ M\subsun~~since the energy density of
the turbulence is an increasing function of the mass of the
cluster. Based on the scenario of hierarchical structure formation,
massive clusters result from a complex merging history, during which
each cluster-cluster collision could have contributed to provide
energy for cosmic ray acceleration.

Finally, as we have seen in Sect. \ref{MiniHalos}, radio mini-halos
have also been observed in clusters. They are located at the centre of
cooling flow clusters and surround a powerful radio galaxy. Similarly
to giant radio halos and relics, the electrons in radio mini-halos
have short radiative lifetimes due to the high magnetic fields present
in cooling cores \citep{Taylor02}. The observed radio emission is thus
not due to the radio lobes of the central AGN. Unlike the giant
sources, mini-halos are typically found in clusters not disturbed by
major mergers (Sect. \ref{MiniHalos}). Again, two possible classes of
models have been proposed. Relativistic electrons could have again an
hadronic origin \citep{Pfrommer04}. Or they could be a relic
population of (primary) relativistic electrons re-accelerated by MHD
turbulence, with the necessary energy supplied by the cooling flow
\citep{Gitti02}. The re-acceleration model by \citet{Gitti02} has been
successfully applied to two cooling flow clusters \citep{Gitti02,
  Gitti04}.  The observed correlation between the mini-halo and
cooling flow power has also given support to a primary origin of the
relativistic electrons \citep{Gitti04, Gitti07a}. However, there also
seems to be some observational and theoretical evidence to support
hadronic origin (\citealp{Kempner04a} and references
therein). Additionally, in two clusters (A~2142:
\citealp{Markevitch00}; RXJ 1347.4$-$1145: \citealp{Gitti07a,
  Gitti07b}), we got indications that cluster mergers and cooling
flows may act simultaneously in powering mini-halo emission in the
rare and peculiar clusters in which they coexist. Further theoretical
and observational studies are indeed essential due to the low number
of known radio mini-halos (Sect. \ref{MiniHalos}).

\section{Measurement of intracluster magnetic fields}
\label{Magnetic}

As stressed above, the presence of diffuse and extended synchrotron
emission in galaxy clusters indicates the existence of weak magnetic
fields in the cluster volume. Different possibilities for their origin
have been proposed which are reviewed by \citet{Dolag08} - Chapter 15,
this volume. Radio observations of galaxy clusters allow us to
measure intracluster magnetic fields and test the different theories
on their origin, as reviewed by \citet{Carilli02} and
\citet{Govoni04}.  In the following the main methods to study magnetic
field intensity and, eventually, structure are summarised.

\subsection{Equipartition magnetic fields}
\label{Equip}
 
In the optically thin case, the total monochromatic emissivity
$J(\nu)$ from a set of relativistic electrons in a magnetic field
$\textit{{\textbf B}}$ depends on a) the magnetic field strength, b)
the energy distribution of the electrons, which is usually assumed to
be a power law (Eq. \ref{eq:ensp}), and c) the pitch angle between the
electron velocity and the magnetic field direction ($\theta$)
\be 
J(\nu) \propto N_0 (B~{\rm sin}\theta)^{1+\alpha} \nu^{-\alpha}, \label{eq:sync}
\ee
where $\alpha = (\delta-1)/2$ is the spectral index of the
synchrotron spectrum\footnote{$\delta$ is the electron energy index,
  see Eq. \ref{eq:ensp}.}.

Synchrotron emission from diffuse and extended radio sources can give
us a direct measure for the intensity of cluster magnetic fields if
the relativistic electron flux is measured or constrained. That can be
achieved, for example, if Compton-produced X-ray (and gamma-ray)
emission was detected simultaneously (see Sect. \ref{IC}). In the case
of polarised radio emission, we can also get an indication of the
projected magnetic field orientation and its degree of ordering. To
break the degeneracy between magnetic field strength and electron
density (Eq. \ref{eq:sync}), and to obtain a measure for cluster
magnetic fields from the observed luminosity of radio sources, it is
typically assumed that the energy density of the relativistic plasma
within a radio source is minimum
\be
U_{\rm tot} = U_{\rm el} + U_{\rm pr} + U_{B} \label{Utot} = U_{\min},
\ee 
where $U_{B}$ is the energy density in magnetic fields, and
$U_{\rm el}$ and $U_{\rm pr}$ are the energy in electrons and in protons
respectively. The energy in the heavy particles (protons) is
considered to be related to the electron energy
\be
U_{\rm pr}= k U_{\rm el}.
\ee 
The value of $k$ depends on the mechanism of
(re-)acceleration of electrons, whose physical details, as seen above,
are still unknown. A typical value of $k$=1 is adopted for halo and
relic sources. Another important assumption of this method relates to
the value of the filling factor, $\Phi$, i.e. the fraction of the
source volume $V$ occupied by magnetic field and relativistic
particles. The energy density in magnetic field is given by
\be
U_{B} = \frac{B^2}{8\pi} \Phi V.
\ee 
It is usually considered that particles and magnetic fields
occupy the entire volume, i.e. $\Phi=1$. It can be derived easily that
the condition of minimum energy is obtained when the contributions of
cosmic rays and magnetic fields is approximately equal
\be
U_B = \frac{3}{4} (1+k) U_{\rm el}.
\ee
This is the so-called classical equipartition assumption,
which allows us to estimate the magnetic field of a radio source from
its radio luminosity $L$ (see \citealp{Pacholczyk70} for a rigorous
derivation)
\be
B_{\rm eq} \propto \left[ \frac{L (1+k)}{\Phi V} \right ]^{2/7} \label{eq:Beq}.
\ee
In the standard approach presented above, $L$ is the
observed synchrotron luminosity between two fixed frequencies $\nu_1$
and $\nu_2$ (usually $\nu_1=$10 MHz and $\nu_2=$100 GHz). In this way,
however, the integration limits are variable in terms of the energy of
the radiating electrons, since, based on Eq. \ref{eq:sync}, electron
energies corresponding to $\nu_1$ and $\nu_2$ depend on magnetic field
values. This point is particularly relevant for the lower limit, owing
to the power-law shape of the electron energy distribution and to the
expected presence of low energy electrons in radio
halos/relics. Alternatively, it has been suggested to derive
equipartition quantities by integrating the electron luminosity over
an energy range ($\gamma_{\min} - \gamma_{\max}$) \citep{Brunetti97,
  Beck05}.  It can be shown that, for $\gamma_{\min} \ll \gamma_{\max}$
and $\alpha > 0.5$, the new expression for the equipartition magnetic
field is
\be
B'_{\rm eq} \sim 1.1 
\gamma_{\min} ^{{\frac{1-2\alpha}{3+\alpha}}}
B_{eq}^{{\frac{7}{2(3+\alpha)}}},  \label{eq:Bpeq}
\ee
where $B_{\rm eq}$ is the equipartition magnetic field expressed
in Gauss derived through Eq. \ref{eq:Beq}. Typically, for $B_{\rm eq} \sim
\mu$G, $\gamma_{\min} \sim 100$ and $\alpha \sim 0.75-1$, this new
approach gives magnetic field values 2 to 5 times larger than the
standard method.

Estimates of equipartition fields on scales as large as $\sim$1 Mpc
give magnetic field intensities in the range 0.1-1 $\mu$G. As we have
seen, these estimates are based on several assumptions both on
different physical properties of the radio emitting region (e.g. the
filling factor $\Phi$ and the ratio between electron and proton
energies $k$), and on the condition of minimum energy of the observed
relativistic plasma.  Since the validity of these assumptions is not
obvious, one has to be aware of the uncertainties and thus of the
limits inherent to the equipartition determination of magnetic fields.

\subsection{Compton scattering of CMB photons}
\label{IC}

As reviewed by \citet{Rephaeli08} - Chapter 5, this volume, 3K
microwave background photons can be subject to Compton scattering by
electrons in the cluster volume.  If the presence of thermal particles
in the ICM results in a distortion of the Cosmic Microwave Background
(CMB) spectrum well known as ``Sunyaev-Zel'dovich effect''
\citep{Sunyaev72}, non-thermal hard X-ray (HXR) photons are produced
via Compton scattering by the same cosmic rays that are responsible
for the synchrotron emission observed at radio wavelengths.  Compton
scattering increases the frequency of the incoming photon through
\be
\nu_{\rm out} = \frac{4}{3} \gamma^2 \nu_{\rm in} \label{eq:IC}.
\ee
The Planck function of the CMB peaks at $\nu_{\rm in} \sim
1.6{\times}10^{11}$ Hz. Based on Eq. \ref{eq:IC}, for typical energies
of relativistic electrons in clusters ($\gamma \sim 1000 - 5000$), the
scattered photons fall in the X-ray and gamma-ray domain ($\sim
2\times10^{17} - 5\times10^{18}$ Hz, i.e. $\sim 0.8 - 20.7$ keV).

Non-thermal HXR emission from galaxy clusters due to Compton
scattering of CMB photons was predicted more than 30 years ago
\citep[e.g.][]{Rephaeli77} and has now been detected in several
systems (\citealp{Rephaeli08} - Chapter 5, this volume;
\citealp{Fusco07} and references therein).  Alternative
interpretations to explain the detected non-thermal X-ray emission
have been proposed in the literature \citep{Blasi99, Ensslin99,
  Blasi00, Dogiel00, Sarazin00}.  However, these hypotheses seem to be
ruled out by energetic considerations, because of the well known
inefficiency of the proposed non-thermal Bremsstrahlung (NTB)
mechanism. NTB emission of keV regime photons with some power $P$
immediately imply about 10$^5$ times larger power to be dissipated in
the plasma that seems to be unrealistic in a quasi-steady model
\citep{Petrosian01, Petrosian03}. For a more detailed treatment of the
origin of HXR emission from galaxy clusters, see the review by
\citet[]{Petrosian08} - Chapter 10, this volume.

The detection of non-thermal HXR and radio emission, produced by the
same population of relativistic electrons, allows us to estimate
unambiguously the volume-averaged intracluster magnetic
field. Following the exact derivations by \citet{Blumenthal70}, the
equations for the synchrotron flux $f_{\rm syn}$ at the frequency $\nu_R$
and the Compton X-ray flux $f_{\rm C}$ at the frequency $\nu_X$ are
\be
f_{\rm syn} (\nu_R) = \frac{\Phi V}{4 \pi {D_{\rm L}}^2} \frac{4 \pi {\rm
e}^3}{(m_{\rm e} {\rm c}^{2})^{\delta}} N_0 B^{\frac{\delta + 1}{2}} \left(
\frac{3 {\rm e}}{4 \pi m_{\rm e} {\rm c}} \right)^{\frac{\delta - 1}{2}} a(\delta) \nu_R^{- \frac{\delta-1}{2}}, \label{eq:Ss}
\ee
\be
f_{\rm C}(\nu_X) = \frac{\Phi V}{4 \pi {D_{\rm L}}^2} \frac{8 \pi^2 {\rm
r}_0^2}{{\rm c}^2} {\rm h}^{-\frac{\delta+3}{2}} N_0 (m_{\rm e} {\rm c}^2)^{(1 -
\delta)} ({\rm k} T)^{\frac{\delta+5}{2}} F({\delta}) \nu_X^{- \frac{\delta-1}{2}}. \label{eq:Sx}
\ee
Here ${\rm h}$ is the Planck constant, $V$ is the volume of the
source and $\Phi$ the filling factor, $D_{\rm L}$ is the luminosity distance
of the source, $B$ the magnetic field strength, $T$ the radiation
temperature of the CMB, ${\rm r}_0$ the classical electron radius (or
Thomson scattering length), $N_0$ and $\delta$ are the amplitude and
the spectral index of the electron energy distribution
(Eq. \ref{eq:ensp}). The values of the functions $a(\delta)$ and
$F(\delta)$ for different values of $\delta$ can be found in
\citet{Blumenthal70}. The field $B$ can thus be estimated directly
from these equations
\be
B \propto \left(\frac{f_{\rm syn} (\nu_R)}{f_{\rm C}(\nu_X)}
\right)^{{\frac{2}{\delta+1}}} \left(
\frac{\nu_R}{\nu_X}\right)^{{\frac{\delta-1}{\delta+1}}}.
\ee
Typical cluster magnetic field values of $\sim 0.1 - 0.3$
$\mu$G are obtained \citep[e.g.][]{Rephaeli99, Fusco99, Fusco00,
  Fusco01, Rephaeli03, Rephaeli06}. Compared to equipartition
measures, this method has the great advantage of using only
observables, assuming only that the spatial factors in the expressions
for the synchrotron and Compton fluxes \citep{Rephaeli79} are
identical.

\subsection{Faraday rotation measure}
\label{RM}

Faraday rotation analysis of radio sources in the background or in the
galaxy clusters themselves is one of the key techniques used to obtain
information on the cluster magnetic fields.  The presence of a
magnetised plasma between an observer and a radio source changes the
properties of the polarised emission from the radio source.  Therefore
information on cluster magnetic fields along the line-of-sight can be
determined, in conjunction with X-ray observations of the hot gas,
through the analysis of the Rotation Measure (RM) of radio sources
\citep[e.g.][]{Burn66}.

The polarised synchrotron radiation coming from radio galaxies
undergoes the following rotation of the plane of polarisation as it
passes through the magnetised and ionised intracluster medium
\be
 \Psi_{\rm Obs}(\lambda)=\Psi_{\rm Int}+\lambda ^2 \times RM,
\label{psi}
\ee
where $\Psi_{\rm Int}$ is the intrinsic polarisation angle, and
$\Psi_{\rm Obs}(\lambda)$ is the polarisation angle observed at a
wavelength $\lambda$.  The RM is related to the thermal electron
density ($n_{\rm e}$), the magnetic field along the line-of-sight
($B_{\|}$), and the path-length ($L$) through the intracluster medium
according to
\be
 RM_{~\rm[rad~m^{-2}]} = 812\int\limits_0^{L_{\rm[kpc]}} n_{\rm e~[cm^{-3}]}
 B_{\|~[{\mu}{\rm G}]} {\rm d}l.
\label{rm}
\ee
Polarised radio galaxies can be mapped at several
frequencies to produce, by fitting Eq.~\ref{psi}, detailed RM images.
Once the contribution of our Galaxy is subtracted, the RM should be
dominated by the contribution of the intracluster medium, and
therefore it can be used to estimate the cluster magnetic field
strength along the line of sight.
 
The RM observed in radio galaxies may not be all due to the cluster
magnetic field if the RM gets locally enhanced by the intracluster
medium compression due to the motion of the radio galaxy
itself. However a statistical RM investigation of point sources
\citep{Clarke01, Clarke04} shows a clear broadening of the RM
distribution toward small projected distances from the cluster centre,
indicating that most of the RM contribution comes from the
intracluster medium. This study included background sources, which
showed similar enhancements as the embedded sources.

We also note that there are inherent uncertainties in the
determination of field values from Faraday Rotation measurements,
stemming largely from the unknown small-scale tangled morphology of
intracluster fields, their large-scale spatial variation across the
cluster, and from the uncertainty in modelling the gas density profile
\citep[see, e.g.,][]{Goldschmidt93, Newman02, Rudnick03a, Ensslin03b,
  Murgia04}.

\begin{figure}    
\centering
\includegraphics[width=0.65\textwidth]{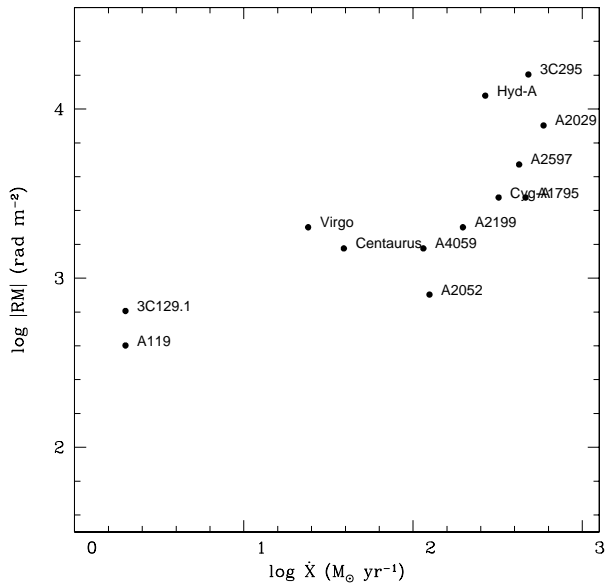}
\caption{RM magnitudes of a sample of radio galaxies located in
    cooling flow clusters, plotted as a function of the cooling flow
    rate \.{X} \citep[from][]{Taylor02}.}
\label{fig:RM}
\end{figure}

RM studies of radio galaxies have been carried out on both statistical
samples \citep[e.g.][]{Lawler82, Vallee86, Kim90, Kim91, Clarke01} and
individual clusters by analysing detailed high resolution RM images
\citep[e.g.][]{Perley91, Taylor93, Feretti95, Feretti99b, Govoni01a,
  Taylor01, Eilek02, Govoni06, Taylor07, Guidetti07}.  Both for interacting and
relaxed (cooling flow) clusters the RM distribution of radio galaxies
is generally patchy, indicating that cluster magnetic fields have
structures on scales as low as 10 kpc or less.  RM data are usually
consistent with central magnetic field strengths of a few $\mu$G. But,
radio galaxies at the centre of relaxed clusters have extreme RM, with
the magnitude of the RM roughly proportional to the cooling flow rate
(see Fig. \ref{fig:RM}). Strong magnetic fields are derived in the
high density cooling-core regions of some clusters, with values
exceeding $\sim 10$ $\mu$G (e.g., in the inner region of Hydra A, a
value of $\sim 35$ $\mu$G was deduced by \citealp{Taylor02}).  It
should be emphasised that such high field values are clearly not
representative of the mean fields in large extended regions.

\citet{Dolag01b} showed that, in the framework of hierarchical cluster
formation, the correlation between two observable parameters, the RM
and the cluster X-ray surface brightness in the source location, is
expected to reflect a correlation between the cluster magnetic field
and gas density.  Therefore, from the analysis of the RM versus X-ray
brightness it is possible to infer the trend of magnetic field versus
gas density.
 
On the basis of the available high quality RM images, increasing
attention is given to the power spectrum of the intracluster magnetic
field fluctuations.  Several studies \citep{Ensslin03a, Murgia04} have
shown that detailed RM images of radio galaxies can be used to infer
not only the cluster magnetic field strength, but also the cluster
magnetic field power spectrum.  The analyses of \citet{Vogt03, Vogt05}
and \citet{Guidetti07} suggest that the power spectrum is of the
Kolmogorov type, if the auto-correlation length of the magnetic
fluctuations is of the order of few kpc. However, \citet{Murgia04} and
\citet{Govoni06} pointed out that shallower magnetic field power
spectra are possible if the magnetic field fluctuations extend out to
several tens of kpc.

\subsection{Comparison of the different methods}
\label{CompB}

As shown in Table 3 of \citet{Govoni04}, the different methods
available to measure intracluster magnetic fields show quite
discrepant results (even more than a factor 10). RM estimates are
about an order of magnitude higher than the measures derived both from
the synchrotron diffuse radio emission and the non-thermal hard X-ray
emission ($\sim 1-5$~$\mu$G vs. $\sim0.2-1$~$\mu$G).

This can be due to several factors. Firstly, equipartition values are
severely affected by the already mentioned physical assumptions of
this method. Secondly, while RM estimates give a weighted average of
the field along the line of sight, equipartition and Compton
scattering measures are made by averaging over larger
volumes. Additionally, discrepancies can be due to spatial profiles of
both the magnetic field and the gas density not being constant all
over the cluster \citep{Goldschmidt93}, or due to compressions,
fluctuations and inhomogeneities in the gas and in the magnetic field,
related to the presence of radio galaxies or to the dynamical history
of the cluster (e.g. on-going merging events) \citep{Beck03,
  Rudnick03b, Johnston04}.  Finally, a proper modelling of the Compton
scattering method should include a) the effects of aged electron
spectra, b) the expected radial profile of the magnetic field, and c)
possible anisotropies in the pitch angle distribution of electrons
\citep{Brunetti01, Petrosian01}.

An additional method of estimating cluster magnetic fields comes from
the X-ray analysis of cold fronts \citep{Vikhlinin01}. These X-ray
cluster features, discovered by \citet{Markevitch00} thanks to the
exquisite spatial resolution of the {\sl Chandra} satellite, result
from dense cool gas moving with near-sonic velocities through the less
dense and hotter ICM. Cold fronts are thus subject to Kelvin-Helmholtz
(K-H) instability that, for typical cluster and cold front properties
(Mach number, gas temperatures, cluster-scale length), could quickly
disturb the front outside a narrow ($\lesssim 10^{\circ}$) sector in
the direction of the cool cloud motion\footnote{For a more precise
  treatment, see \citet{Vikhlinin01} and
  \citet{Markevitch07}.}. Through the {\sl Chandra} observation of
A~3667, \citet{Vikhlinin01} instead revealed a cold front that is
stable within a $\pm 30^{\circ}$ sector. They showed that a 
$\sim 10$~$\mu$G magnetic field oriented nearly parallel to the front is able
to suppress K-H instability, thus preserving the front structure, in a
$\pm 30^{\circ}$ sector. The estimated magnetic field value,
significantly higher than the typical measures given by the other
methods outside cluster cooling flows, is likely an upper limit of the
absolute field strength. Near the cold front the field is actually
amplified by tangential gas motions \citep[see][]{Vikhlinin01}.

Variations of the magnetic field structure and strength with the
cluster radius have been recently pointed out by \citet{Govoni06}.  By
combining detailed multi-wavelength and numerical studies we will get
more insight into the strength and structure of intracluster magnetic
fields, and into their connection with the thermodynamical evolution
of galaxy clusters. More detailed comparisons of the different
approaches for measuring intracluster magnetic fields can be found,
for instance, in \citet{Petrosian03} and \citet{Govoni04}.

\section{Diffuse radio emission in galaxy clusters: open questions and 
perspectives}
\label{Discussion}

Significant progress has been made recently in our knowledge on the
non-thermal component of galaxy clusters. A number of open questions
arise in assessing the current theoretical and observational status.

First of all, we need to test the current theories on the origin of
the large-scale non-thermal component in clusters (magnetic field and
cosmic rays). If at present primary models seem to be the favourite
acceleration mechanisms for intracluster electrons, secondary models
cannot be ruled out. Among other things, it will be necessary to
establish: How common is the non-thermal component in clusters? Is it
really hosted {\sl only} in merging systems (as present observational
results suggest) or do {\sl all} clusters have a radio halo/relic? If
this latter hypothesis is correct, how should we modify the radio
power versus X-ray luminosity correlation (Sect. \ref{Halos})? If
shocks and turbulence related to cluster mergers are instead the
mechanisms responsible for electron re-acceleration, why have extended
radio sources not been detected in {\sl all} merging clusters? Is this
related to other physical effects (i.e. the merging event alone is not
enough to produce intracluster cosmic rays), as the correlation
between radio power and cluster mass seems to indicate, or it is due
to a lack of sensitivity of the current instruments (i.e. all merging
clusters host radio halos and relics, but a large fraction of these
sources lies below the sensitivity limit of present telescopes)? 

Among the previous questions, the most difficult to answer {\sl at
  present} are those that involve the study of low-luminous X-ray
clusters, for which the limits of current radio observations are
particularly severe. By extrapolating to low radio and X-ray
luminosity the $P_{\nu}$ - $L_X$ relation (Sect. \ref{Halos}),
\citet{Feretti07} have estimated that, if present, halos with typical
sizes of 1 Mpc in intermediate/low-luminous X-ray clusters
($L_{X[0.1-2.4~{\rm keV}]} \lesssim 5 \times
10^{44}~{h_{70}}^{-2}~{\rm erg}~{\rm s}^{-1}$)\footnote{Converted from
  the bolometric X-ray luminosity limit in \citet{Feretti07} to the
  $0.1-2.4$ keV band luminosity using Table 5 of \citet{Boehringer04}
  and assuming typical ICM temperature values ($T_X \sim 5-10$~keV).}
would actually have a radio surface brightness lower than the current
limits obtained in the literature and in the NVSS. At higher X-ray
luminosities, more constraints on radio halo statistics have recently
been obtained by \citet{Venturi07} and \citet{Brunetti07}. They
carried out GMRT\footnote{The Giant Metrewave Radio Telescope (GMRT)
  is operated by the National Centre for Radio Astrophysics of the
  Tata Institute of Fundamental Research (NCRA-TIFR).}  observations
at 610 MHz of 34 luminous ($L_{X[0.1-2.4~{\rm keV}]} \gtrsim 5 \times
10^{44}~{h_{70}}^{-2}~{\rm erg}~{\rm s}^{-1}$) clusters with 0.2
$\lesssim z \lesssim$ 0.4. The bulk of the galaxy clusters in their
sample does not show any diffuse central radio emission, with radio
luminosity upper limits that are well below the $P_{\nu}$ - $L_X$
relation derived from the previously known radio halos. The net
bimodality of the cluster distribution in the $P_{\nu}$ - $L_X$ plane
support primary models against secondary models. Actually, the former
predict a relatively fast ($\approx 10^8$ yrs) transition of clusters
from a radio quiet state to the observed $P_{\nu}$ - $L_X$
correlation, where they remain for $\lesssim$ 1 Gyr. A significantly
wider scatter around the $P_{\nu}$ - $L_X$ correlation is instead
expected in the frame of secondary models, that could be reconciled
with observations only assuming the existence of strong dissipation of
magnetic fields in clusters \citep[see][and references
  therein]{Brunetti07}.

On smaller scales, a larger sample of radio mini-halos is required to
test the current theories on the origin of radio emission in this
class of sources \citep{Gitti02, Pfrommer04}. Recent results suggest
that cooling flows and mergers could act simultaneously, when they
co-exist, in providing energy to the relic population of relativistic
electrons injected into the ICM by AGNs, thus powering mini-halo radio
emission \citep{Gitti07a}.

As discussed in Sect. \ref{Magnetic}, radio observations of galaxy
clusters offer a unique tool to estimate strength and structure of
large-scale magnetic fields, allowing to test the different scenarios
of their origin. Several observational results show that magnetic
fields of the order of $\sim \mu$G are common in clusters. Through
combined numerical and observational analyses, \citet{Murgia04} and
\citet{Govoni06} have shown that detailed morphology and polarisation
information of radio halos may provide important constraints on the
strength and structure of intracluster magnetic fields. However,
discrepant results have been obtained up to now (Sect. \ref{CompB})
and more detailed information on magnetic fields is still needed.

A better knowledge of the physics of the non-thermal component in
galaxy clusters will have important cosmological implications. If it
will be confirmed that the presence of giant halos and relics is
related to cluster mergers, the statistical properties of these radio
sources will allow us to test the current cluster formation scenario,
giving important hints on large-scale structure formation and, thus,
cosmological parameters \citep[e.g.][]{Evrard02}.

Additionally, we will be able to estimate how the gravitational energy
released during cluster mergers is redistributed between the thermal
ICM and the relativistic plasma \citep[e.g.][]{Sarazin05}. The effects
of magnetic fields on the thermodynamical evolution of large-scale
structures will be evaluated, as well as the contribution of the
non-thermal pressure to the estimate of mass and temperature in galaxy
clusters \citep[e.g.][]{Dolag00b, Dolag01a, Colafrancesco04}. Cluster
scaling laws, such as mass vs. temperature, are actually key
ingredients to derive cosmological constraints from galaxy clusters
\citep[e.g.][]{Ettori04, Arnaud05}.

Finally, a better knowledge of extended radio sources in clusters is
indeed essential for complementary cosmological studies, e.g. the
epoch of re-ionisation (EoR). It has been proven that radio halos and
relics are the strongest extra-galactic foreground sources to be
removed in order to probe the EoR through the study of the redshifted
21 cm emission from neutral hydrogen
\citep[e.g.][]{DiMatteo04}. Better models for the diffuse radio
emission have to be inserted into numerical simulations of the EoR 21
cm emission, in order to understand how to remove efficiently the
contamination due to radio halos and relics.

\begin{figure}    
\centering
\includegraphics[width=0.75\textwidth]{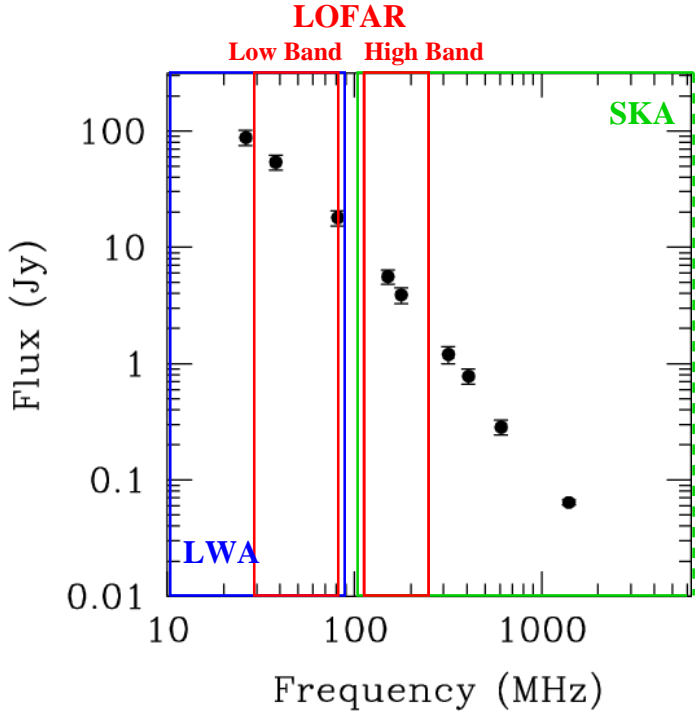}
\caption{Spectrum of the diffuse radio source in A~1914
    \citep[from][]{Bacchi03}. Superimposed the frequency range covered
    by LWA (10 MHz -- 88 MHz, in blue), LOFAR (Low Band: 30 MHz -- 80
    MHz, High Band: 110 MHz -- 240 MHz, in red) and SKA (100 MHz -- 25
    GHz, in green). The low-frequency domain covered by the next
    generation radio-telescopes is optimal for the detection of high
    spectral index radio sources, such as radio halos, mini-halos and
    relics.}
\label{fig:spectrum-lowf}  
\end{figure}

An increase in the number of known radio halos/relics, as well as
higher resolution and sensitivity observations, are essential to
answer the main open questions, summarised at the beginning of this
section, about the nature of diffuse radio emission in clusters. As
shown in Fig. \ref{fig:spectrum-lowf}, halos and relics are difficult
to detect in the GHz range due to their steep spectra. Several
observations performed with the currently available low-frequency
instruments (e.g. GMRT: \citealp{Venturi07}; VLA: \citealp{Kassim01,
  Orru07}) confirm the interest in studying this class of radio
sources at high wavelengths. A short term perspective in the study of
radio halos and relics is thus to fully exploit those instruments that
are already available for observations in the MHz range of the
electromagnetic spectrum with good enough sensitivity (approximately
from some tens of $\mu$Jy/beam to some mJy/beam) and angular
resolution (roughly some tens of arcsec). However, in order to make a
proper comparison between observational results and current
theoretical models about the origin of radio halos and relics, we need
multi-frequency observations of {\sl statistical} samples of diffuse
radio sources. Current telescopes require too long exposure-time per
cluster ($\gtrsim 1-2$ hours) to reach the sensitivity limits
necessary for detecting radio halos/relics, making statistical
analyses of diffuse radio emission in clusters extremely
time-demanding.

The low-frequency range covered by a new generation of radio
telescopes (Long Wavelength Array - LWA; Low Frequency Array - LOFAR;
Square Kilometre Array - SKA), together with their gain in sensitivity
and resolution, will increase dramatically the statistics on the
number of known radio halos and relics.  Not only these instruments
will cover the optimal frequency range for halo/relic detection (see
Fig. \ref{fig:spectrum-lowf}), but also their gain in sensitivity and
resolution will be of the order of 10 to 1000 \citep[see Table 2
  of][]{Brueggen05}, allowing observations of statistical samples of
diffuse and extended radio sources. A LOFAR survey at 120 MHz,
covering half the sky to a 5 $\sigma$ flux limit of 0.1 mJy (1 hour
integration time per pointing), could detect $\sim$1000 halos/relics,
of which 25~\% at redshift larger than $z \sim$ 0.3
\citep{Roettgering03}. \citet{Feretti04b} have estimated that, with 1
hour integration time at 1.4 GHz, 50~\% of the SKA collecting area will
allow us to detect halos and relics of total flux down to 1 mJy at any
redshift, and down to 0.1 mJy at high redshift. Based on our current
knowledge, more than three (fifteen) hundred diffuse cluster radio
sources are expected at 1.5 GHz on the full sky at the 1 mJy (0.1 mJy)
flux limit \citep{Ensslin02}.

With statistical samples of halos and relics over a wide redshift
range, we will be able to a) test the correlation between the
non-thermal component and the physical properties of clusters
(dynamical state, mass, X-ray luminosity and temperature...), b)
analyse the redshift evolution of halos and relics, with the
advantages for cosmological studies stressed above, and c) fill the
gap in our knowledge of the last phases of radio galaxy evolution in
clusters (see Sect. \ref{Relics}). Particularly interesting will be
the study of possible presence of non-thermal radio emission at $z
\sim$ 1, i.e. the epoch of formation of the massive galaxy clusters
observed in the local Universe.

The excellent sensitivity, high angular resolution and large number of
spectral channels of the next generation instruments, together with
new techniques of RM synthesis \citep{Brentjens05}, will allow
polarisation mapping and RM studies of radio emission in clusters,
significantly improving our estimates of large-scale magnetic fields.

Future radio observations of galaxy clusters, combined with the new
generation instruments at other wavelengths\footnote{See also
  \citet{Paerels08} - Chapter 19, this volume.} (e.g. sub-mm: ALMA;
X-ray: XEUS, Simbol-X; gamma-rays: GLAST, H.E.S.S., MAGIC\footnote{A
  new generation of ground based imaging Cherenkov telescopes will be
  available soon. The measurements above 10 TeV are crucial to
  distinguish between the Compton scattering and hadronic origins of
  gamma-ray emission from clusters of galaxies. }; ...), will allow us
to open a new window in cosmological studies.

\begin{acknowledgements}
The authors thank ISSI (Bern) for support of the team ``Non-virialized
X-ray components in clusters of galaxies''. CF and FG warmly thank
Luigina Feretti and Matteo Murgia for many useful discussions on the
subject of this paper. CF and SS acknowledge financial support by the
Austrian Science Foundation (FWF) through grants P18523-N16 and
P19300-N16, by the Tiroler Wissenschaftsfonds and through the
UniInfrastrukturprogramm 2005/06 by the BMWF. FG acknowledges
financial support through Grant ASI-INAF I/088/06/0 - High Energy
Astrophysics.
\end{acknowledgements}


\begin{thebibliography}{}

\bibitem[\protect\citeauthoryear{Allen et al.}{2001}]{Allen01} 
Allen, S.W., Ettori, S., \& Fabian, A.C., \mnras{2001}{324}{877}

\bibitem[\protect\citeauthoryear{Arnaud et al.}{2005}]{Arnaud05}
  Arnaud, M., Pointecouteau, E., \& Pratt, G.W., \aeta{2005}{441}{893}

\bibitem[\protect\citeauthoryear{Bacchi et al.}{2003}]{Bacchi03}
Bacchi, M., Feretti, L., Giovannini, G., \& Govoni, F.,
\aeta{2003}{400}{465}

\bibitem[\protect\citeauthoryear{Bagchi et al.}{2002}]{Bagchi02}
Bagchi, J. En{\ss}lin, T., Miniati, F., et al., 2002, New Astron., 7, 249

\bibitem[\protect\citeauthoryear{Bagchi et al.}{2006}]{Bagchi06} 
Bagchi, J., Durret, F., Lima Neto, G.B., \& Paul, S., 2006, Science, 314, 791

\bibitem[\protect\citeauthoryear{Baum \& O'Dea}{1991}]{Baum91} 
Baum, S.A., \& O'Dea, C.P., \mnras{1991}{250}{737}

\bibitem[\protect\citeauthoryear{Beck et al.}{2003}]{Beck03} 
Beck, R., Sukurov, A, Sokoloff, D., \& Wielebinski, R., \aeta{2003}{411}{99}

\bibitem[\protect\citeauthoryear{Beck \& Krause}{2005}]{Beck05} 
Beck, R., \& Krause, M., 2005,  Astron. Nachr., 326, 414 

\bibitem[\protect\citeauthoryear{Berezinsky et al.}{1997}]{Berezinsky97} 
Berezinsky, V.S., Blasi, P., \& Ptuskin, V.S., \aspj{1997}{487}{529}

\bibitem[\protect\citeauthoryear{Blasi \& Colafrancesco}{1999}]{Blasi99} 
Blasi, P. \& Colafrancesco, S., 1999, Astrop. Phys., 12, 169

\bibitem[\protect\citeauthoryear{Blasi}{2000}]{Blasi00}
  Blasi, P., \aspj{2000}{532}{L9}

\bibitem[\protect\citeauthoryear{Blasi}{2004}]{Blasi04}
Blasi, P., 2004, JKAS, 37, 483

\bibitem[\protect\citeauthoryear{Blasi et al.}{2007}]{Blasi07}
  Blasi, P., Gabici, S., \& Brunetti, G., 2007, Int. J. Mod. Phys., A22, 681

\bibitem[\protect\citeauthoryear{Blumenthal \& Gould}{1970}]{Blumenthal70} 
Blumenthal, G.R., \& Gould, R.J., 1970, Rev. Mod. Phys., 42, 237

\bibitem[\protect\citeauthoryear{B\"ohringer et al.}{2004}]{Boehringer04} 
B\"ohringer, H., Schuecker, P., Guzzo, L., et al., \aeta{2004}{425}{367}

\bibitem[\protect\citeauthoryear{Brentjens \& de Bruyn}{2005}]{Brentjens05} 
Brentjens, M.A., \& de Bruyn, A.G., \aeta{2005}{441}{1217}

\bibitem[\protect\citeauthoryear{Br{\"u}ggen et al.}{2005}]{Brueggen05}
  Br\"uggen, M., Falcke, H., Beck, R., et al., 2005,
  ``German LOFAR white paper'', available electronically at
  http://lofar.mpa-garching.mpg.de/glow.html

\bibitem[\protect\citeauthoryear{Brunetti et al.}{1997}]{Brunetti97} 
Brunetti, G., Setti, G., \& Comastri, A., \aeta{1997}{325}{898}

\bibitem[\protect\citeauthoryear{Brunetti et al.}{2001}]{Brunetti01} 
Brunetti, G., Setti, G., Feretti, L., \& Giovannini, G., \mnras{2001}{320}{365}

\bibitem[\protect\citeauthoryear{Brunetti}{2004}]{Brunetti04a}
  Brunetti, G., 2004, JKAS, 37, 493

\bibitem[\protect\citeauthoryear{Brunetti et al.}{2004}]{Brunetti04b}
  Brunetti, G., Blasi, P., Cassano, R., \& Gabici, S., \mnras{2004}{350}{1174}

\bibitem[\protect\citeauthoryear{Brunetti et al.}{2007}]{Brunetti07}
  Brunetti, G., Venturi, T., Dallacasa, D., et al., \aspj{2007}{670}{L5}

\bibitem[\protect\citeauthoryear{Buote \& Tsai}{1996}]{Buote96} 
Buote, D.A., \& Tsai, J.C., \aspj{1996}{458}{27}

\bibitem[\protect\citeauthoryear{Buote}{2001}]{Buote01}
Buote, D.A., \aspj{2001}{553}{L15}

\bibitem[\protect\citeauthoryear{Buote}{2002}]{Buote02} 
Buote, D.A., 2002, in ``Merging processes in galaxy
clusters'', L. Feretti, I.M. Gioia \& G. Giovannini (eds.),
Astrophys. Sp. Sc. Lib. (Kluwer), 272, 79

\bibitem[\protect\citeauthoryear{Burn}{1966}]{Burn66}
Burn, B. J., \mnras{1966}{133}{67}

\bibitem[\protect\citeauthoryear{Burns et al.}{1992}]{Burns92}
Burns, J.O., Sulkanen, M.E., Gisler, G.R., \& Perley, R.A., \aspj{1992}{388}{L49}

\bibitem[\protect\citeauthoryear{Bykov et al.}{2000}]{Bykov00}
Bykov, A.M., Bloemen, H., \& Uvarov, Yu.A., \aeta{2000}{362}{886}

\bibitem[\protect\citeauthoryear{Bykov et al.}{2008}]{Bykov08}
Bykov, A.M., Dolag, K., \& Durret, F., 2008, SSR, in press

\bibitem[\protect\citeauthoryear{Carilli \& Taylor}{2002}]{Carilli02}
  Carilli, C.L., \& Taylor, G.B., \araa{2002}{40}{319}

\bibitem[\protect\citeauthoryear{Cassano \& Brunetti}{2005}]{Cassano05} 
Cassano, R., \& Brunetti, G., \mnras{2005}{357}{1313}

\bibitem[\protect\citeauthoryear{Cassano et al.}{2006}]{Cassano06} 
Cassano, R., Brunetti, G., \& Setti, G., \mnras{2006}{369}{1577}

\bibitem[\protect\citeauthoryear{Cassano et al.}{2007}]{Cassano07} 
Cassano, R., Brunetti, G., Setti, G., Govoni, F., \& Dolag, K., 
\mnras{2007}{378}{1565}

\bibitem[\protect\citeauthoryear{Clarke et al.}{2001}]{Clarke01}
Clarke, T.E., Kronberg, P.P., \& B{\" o}hringer, H., \aspj{2001}{547}{L111}

\bibitem[\protect\citeauthoryear{Clarke}{2004}]{Clarke04}
Clarke, T.E., 2004, JKAS, 37, 337

\bibitem[\protect\citeauthoryear{Clarke \&
    En{\ss}lin}{2006}]{Clarke06} Clarke, T.E., \& En{\ss}lin, T.A., 
\aj{2006}{131}{2900}

\bibitem[\protect\citeauthoryear{Colafrancesco}{1999}]{Colafrancesco99} 
Colafrancesco, S., 1999, in ``Diffuse thermal and relativistic plasma
in galaxy clusters'', H. B\"ohringer, L. Feretti \& P. Schuecker
(eds.),  MPE Rep., 271, 269 (astro-ph/9907329) 

\bibitem[\protect\citeauthoryear{Colafrancesco et al.}{2004}]{Colafrancesco04} 
Colafrancesco, S., Dar, A., \& De R\'ujula, A., \aeta{2004}{413}{441}

\bibitem[\protect\citeauthoryear{Condon et al.}{1998}]{Condon98} 
Condon, J.J., Cotton, W.D., Greisen, E.W., et al., \aj{1998}{115}{1693}

\bibitem[\protect\citeauthoryear{Deiss et al.}{1997}]{Deiss97} 
Deiss, B.M., Reich, W., Lesch, H., \& Wielebinski, R., \aeta{1997}{321}{55}

\bibitem[\protect\citeauthoryear{Delain \& Rudnick}{2006}]{Delain06}
Delain, K.M., \& Rudnick, L., 2006,  Astron. Nachr., 327, 561

\bibitem[\protect\citeauthoryear{Dennison}{1980}]{Dennison80}
 Dennison, B., \aspj{1980}{239}{L93}

\bibitem[\protect\citeauthoryear{Di Matteo et al.}{2004}]{DiMatteo04} 
Di Matteo, T., Ciardi, B., \& Miniati, F., \mnras{2004}{355}{1053}

\bibitem[\protect\citeauthoryear{Dogiel}{2000}]{Dogiel00} 
Dogiel, V.A., \aeta{2000}{357}{66}

\bibitem[\protect\citeauthoryear{Dolag \& En{\ss}lin}{2000}]{Dolag00a} 
Dolag, K., \& En{\ss}lin, T.A, \aeta{2000}{362}{151}

\bibitem[\protect\citeauthoryear{Dolag \& Schindler}{2000}]{Dolag00b} 
Dolag, K., \& Schindler, S., \aeta{2000}{364}{491}

\bibitem[\protect\citeauthoryear{Dolag et al.}{2001a}]{Dolag01a} 
Dolag, K., Evrard, A., \& Bartelmann, M., \aeta{2001a}{369}{36}

\bibitem[\protect\citeauthoryear{Dolag et al.}{2001b}]{Dolag01b} 
Dolag, K., Schindler, S., Govoni, F., \& Feretti, L., \aeta{2001b}{378}{777}

\bibitem[\protect\citeauthoryear{Dolag et al.}{2008}]{Dolag08} 
Dolag, K., Bykov, A.M., \& Diaferio, A., 2008, SSR, in press

\bibitem[\protect\citeauthoryear{Eilek \& Owen}{2002}]{Eilek02} Eilek,
  J.A., \& Owen, F.N., \aspj{2002}{567}{202}

\bibitem[\protect\citeauthoryear{En{\ss}lin et al.}{1998}]{Ensslin98}
  En{\ss}lin, T.A., Biermann, P.L., Klein, U., \& Kohle, S.,
  \aeta{1998}{332}{395}

\bibitem[\protect\citeauthoryear{En{\ss}lin et al.}{1999}]{Ensslin99}
  En{\ss}lin, T.A., Lieu, R., \& Biermann, P.L., \aeta{1999}{344}{409}

\bibitem[\protect\citeauthoryear{En{\ss}lin \& Gopal-Krishna}{2001}]{Ensslin01} 
En{\ss}lin, T.A., \& Gopal-Krishna, \aeta{2001}{366}{26}

\bibitem[\protect\citeauthoryear{En{\ss}lin \& R\"ottgering}{2002}]{Ensslin02}
  En{\ss}lin, T.A., \& R\"ottgering, H.J.A., \aeta{2002}{396}{83}

\bibitem[\protect\citeauthoryear{En{\ss}lin \&
    Vogt}{2003}]{Ensslin03a} En{\ss}lin T.A., \& Vogt C., \aeta{2003}{401}{835}

\bibitem[\protect\citeauthoryear{En{\ss}lin et al.}{2003}]{Ensslin03b} 
En{\ss}lin T.A., Vogt C., Clarke, T.E., \& Taylor, G.B., \aspj{2003}{597}{870}

\bibitem[\protect\citeauthoryear{Ettori et al.}{1998}]{Ettori98} 
Ettori, S., Fabian, A.C., \& White, D.A., \mnras{1998}{300}{837}

\bibitem[\protect\citeauthoryear{Ettori et al.}{2004}]{Ettori04} 
Ettori, S., Tozzi, P., Borgani, S., \& Rosati, P., \aeta{2004}{417}{13}

\bibitem[\protect\citeauthoryear{Evrard \& Gioia}{2002}]{Evrard02}
  Evrard, A.E., \& Gioia, I.M., 2002, in ``Merging processes in galaxy
  clusters'', L. Feretti, I.M. Gioia \& G. Giovannini (eds.),
  Astrophys. Sp. Sc. Lib. (Kluwer),
  272, 253

\bibitem[\protect\citeauthoryear{Fabian et al.}{2000}]{Fabian00} 
Fabian, A.C., Sanders, J.S., Ettori, S., et al., \mnras{2000}{318}{65}

\bibitem[\protect\citeauthoryear{Fabian et al.}{2002}]{Fabian02}
  Fabian, A.C., Celotti, A., Blundell, K.M., Kassim, N.E., \& Perley,
  R.A., \mnras{2002}{331}{369}

\bibitem[\protect\citeauthoryear{Feretti et al.}{1995}]{Feretti95}
Feretti, L., Dallacasa, D., Giovannini, G., \& Tagliani, A.,
\aeta{1995}{302}{680}

\bibitem[\protect\citeauthoryear{Feretti et al.}{1997}]{Feretti97}
  Feretti, L., Giovannini, G., \& B\"ohringer, H., 1997,  New Astron., 2, 501

\bibitem[\protect\citeauthoryear{Feretti}{1999}]{Feretti99a} 
Feretti, L., 1999, in ``Diffuse thermal and relativistic plasma
in galaxy clusters'', H. B\"ohringer, L. Feretti \& P. Schuecker
(eds.),  MPE Rep., 271, 3

\bibitem[\protect\citeauthoryear{Feretti et al.}{1999}]{Feretti99b}
 Feretti, L., Dallacasa, D., Govoni, F., et al., \aeta{1999}{344}{472}

\bibitem[\protect\citeauthoryear{Feretti et al.}{2001}]{Feretti01} 
Feretti, L., Fusco-Femiano, R., Giovannini, G., \& Govoni, F.,
\aeta{2001}{373}{106}

\bibitem[\protect\citeauthoryear{Feretti \& Venturi}{2002}]{Feretti02a} 
Feretti, L., \& Venturi, T., 2002, in ``Merging processes in galaxy
clusters'', L. Feretti, I.M. Gioia \& G. Giovannini (eds.),
Astrophys. Sp. Sc. Lib. (Kluwer),
272, 163

\bibitem[\protect\citeauthoryear{Feretti}{2002}]{Feretti02b} 
Feretti, L., 2002, in ``The Universe at low radio frequencies'', Pramesh
Rao, G. Swarup \& Gopal-Krishna (eds.), IAU Symp., 199, 133

\bibitem[\protect\citeauthoryear{Feretti et al.}{2004a}]{Feretti04a}
Feretti, L., Orr\`u, E., Brunetti, G., et al., \aeta{2004a}{423}{111}

\bibitem[\protect\citeauthoryear{Feretti et al.}{2004b}]{Feretti04b}
Feretti, L., Burigana, C., \& En{\ss}lin, T.A., 2004b, "Science with the
Square Kilometre Array", C. Carilli \& S. Rawlings (eds.), New
  Astron. Rev., 48, 1137

\bibitem[\protect\citeauthoryear{Feretti}{2005}]{Feretti05}
Feretti, L., \adv{2005}{36}{729}  

\bibitem[\protect\citeauthoryear{Feretti \& Neumann}{2006}]{Feretti06}
  Feretti, L., \& Neumann, D.M., \aeta{2006}{450}{L21}

\bibitem[\protect\citeauthoryear{Feretti \& Giovannini}{2007}]{Feretti07} 
Feretti, L., \& Giovannini, G., 2007, to be published in ``Panchromatic
view of clusters of galaxies and the large-scale structure'',
M. Plionis, O. Lopez-Cruz \& D. Hughes (eds.) (astro-ph/0703494)

\bibitem[\protect\citeauthoryear{Ferrari}{2003}]{Ferrari03a} 
Ferrari, C., 2003, ``Multi-wavelength analysis of merging galaxy
clusters'', PhD Thesis, Nice, available electronically at 
http://tel.archives-ouvertes.fr/docs/00/04/85/55/PDF/tel-00010416.pdf

\bibitem[\protect\citeauthoryear{Ferrari et al.}{2003}]{Ferrari03b}
Ferrari, C., Maurogordato, S., Cappi, A., \& Benoist, C., \aeta{2003}{399}{813}

\bibitem[\protect\citeauthoryear{Ferrari et al.}{2005}]{Ferrari05}
Ferrari, C., Benoist, C., Maurogordato, S., Cappi, A., \& Slezak, E.,
\aeta{2005}{430}{19}

\bibitem[\protect\citeauthoryear{Ferrari et al.}{2006a}]{Ferrari06a}
  Ferrari, C., Arnaud, M., Ettori, S., Maurogordato, S., \& Rho, J.,
  \aeta{2006a}{446}{417}

\bibitem[\protect\citeauthoryear{Ferrari et al.}{2006b}]{Ferrari06b} 
Ferrari, C., Hunstead, R.W., Feretti, L., Maurogordato, S., \& Schindler,
S., \aeta{2006b}{457}{21}

\bibitem[\protect\citeauthoryear{Finoguenov et al.}{2005}]{Finoguenov05} 
Finoguenov, A., B\"ohringer, H., \& Zhang, Y.-Y., \aeta{2005}{442}{827}

\bibitem[\protect\citeauthoryear{Furusho et al.}{2001}]{Furusho01}
Furusho, T., Yamasaki, N.Y., Ohashi, T., Shibata, R., \& Ezawa, H., 
\aspj{2001}{561}{L165}


\bibitem[\protect\citeauthoryear{Fusco-Femiano et al.}{1999}]{Fusco99}
  Fusco-Femiano, R., Dal Fiume, D., Feretti, L., et al., \aspj{1999}{513}{L21}

\bibitem[\protect\citeauthoryear{Fusco-Femiano et al.}{2000}]{Fusco00}
  Fusco-Femiano, R., Dal Fiume, D., De Grandi, S., et al., 
\aspj{2000}{534}{L7}

\bibitem[\protect\citeauthoryear{Fusco-Femiano et al.}{2001}]{Fusco01}
  Fusco-Femiano, R., Dal Fiume, D., Orlandini, M., et al., \aspj{2001}{552}{L97}

\bibitem[\protect\citeauthoryear{Fusco-Femiano et al.}{2007}]{Fusco07}
  Fusco-Femiano, R., Landi, R., \& Orlandini, M., \aspj{2007}{654}{9}

\bibitem[\protect\citeauthoryear{Gabici \& Blasi}{2003}]{Gabici03}
Gabici, S., \& Blasi, P., \aspj{2003}{583}{695}

\bibitem[\protect\citeauthoryear{Gavazzi et al.}{2003}]{Gavazzi03}
Gavazzi, G., Cortese, L., Boselli, A. et al., \aspj{2003}{597}{210}

\bibitem[\protect\citeauthoryear{Giacintucci et al.}{2005}]{Giacintucci05}
Giacintucci, S., Venturi, T., Brunetti, G., et al., \aeta{2005}{440}{867}

\bibitem[\protect\citeauthoryear{Giacintucci et al.}{2006}]{Giacintucci06}
Giacintucci, S., Venturi, T., Bardelli, S., et al., 2006, New Astron., 11, 437

\bibitem[\protect\citeauthoryear{Giovannini et al.}{1993}]{Giovannini93}
Giovannini, G., Feretti, L., Venturi, T., Kim, K.-T., \& Kronberg, P.P.,
\aspj{1993}{406}{399}

\bibitem[\protect\citeauthoryear{Giovannini et al.}{1999}]{Giovannini99}
Giovannini, G., Tordi, M., \& Feretti, L., 1999,  New Astron., 4, 141

\bibitem[\protect\citeauthoryear{Giovannini \&
    Feretti}{2000}]{Giovannini00} Giovannini, G., \& Feretti, L., 2000,
   New Astron., 5, 335

\bibitem[\protect\citeauthoryear{Giovannini \&
    Feretti}{2002}]{Giovannini02} Giovannini, G., \& Feretti, L., 2002,
  in ``Merging processes in galaxy clusters'', L. Feretti, I.M. Gioia,
  \& G. Giovannini (eds.),  Astrophys. Sp. Sc. Lib. (Kluwer), 272, 197

\bibitem[\protect\citeauthoryear{Girardi \& Biviano}{2002}]{Girardi02}
  Girardi, M., \& Biviano, A., 2002, in ``Merging processes in galaxy
  clusters'', L. Feretti, I.M. Gioia \& G. Giovannini (eds.),
  Astrophys. Sp. Sc. Lib. (Kluwer),
  272, 39

\bibitem[\protect\citeauthoryear{Gitti et al.}{2002}]{Gitti02}
Gitti, M., Brunetti, G., \& Setti, G., \aeta{2002}{386}{456}

\bibitem[\protect\citeauthoryear{Gitti et al.}{2004}]{Gitti04}
Gitti, M., Brunetti, G., Feretti, L., \& Setti, G., \aeta{2004}{417}{1}

\bibitem[\protect\citeauthoryear{Gitti et al.}{2007a}]{Gitti07a}
Gitti, M., Ferrari, C., Domainko, W., Feretti, L., \& Schindler, S., 
\aeta{2007a}{470}{L25}

\bibitem[\protect\citeauthoryear{Gitti et al.}{2007b}]{Gitti07b}
Gitti, M., Piffaretti, R., \& Schindler, S., \aeta{2007b}{472}{383}

\bibitem[\protect\citeauthoryear{Goldshmidt \&
    Rephaeli}{1993}]{Goldschmidt93}
Goldshmidt, O., \& Rephaeli, Y., \aspj{1993}{411}{518}

\bibitem[\protect\citeauthoryear{Gomez et al.}{2002}]{Gomez02} 
G\'omez, P.L., Loken, C., R\"ottiger, K., \& Burns, J.O., \aspj{2002}{569}{122}

\bibitem[\protect\citeauthoryear{Govoni et al.}{2001a}]{Govoni01a}
Govoni, F., Taylor, G.B., Dallacasa, D., et al., \aeta{2001a}{369}{441}

\bibitem[\protect\citeauthoryear{Govoni et al.}{2001b}]{Govoni01b}
Govoni, F., Feretti, L., Giovannini, G., et al., \aeta{2001b}{376}{803}

\bibitem[\protect\citeauthoryear{Govoni et al.}{2001c}]{Govoni01c}
  Govoni, F., En{\ss}lin, T.A., Feretti, L., \& Giovannini, G.,
\aeta{2001c}{379}{807}

\bibitem[\protect\citeauthoryear{Govoni \& Feretti}{2004}]{Govoni04}
  Govoni, F., \& Feretti, L., 2004,  Internl. J. Modern Phys. D, 13, 1549

\bibitem[\protect\citeauthoryear{Govoni et al.}{2005}]{Govoni05}
  Govoni, F., Murgia, M., Feretti, L., et al., \aeta{2005}{430}{L5}

\bibitem[\protect\citeauthoryear{Govoni et al.}{2006}]{Govoni06}
  Govoni, F., Murgia, M., Feretti, L., et al., \aeta{2006}{460}{425}

\bibitem[\protect\citeauthoryear{Guidetti et al.}{2007}]{Guidetti07}
  Guidetti, D., Murgia, M., Govoni, F., et al., 2007, A\&A, in press,
  arXiv:0709.2652

\bibitem[\protect\citeauthoryear{Hoeft \& Br{\"u}ggen}{2007}]{Hoeft07}
  Hoeft, M., \& Br{\"u}ggen, M., \mnras{2007}{375}{77}

\bibitem[\protect\citeauthoryear{Jaffe}{1977}]{Jaffe77} 
Jaffe, W.J., \aspj{1977}{212}{1}

\bibitem[\protect\citeauthoryear{Johnston-Hollitt et al.}{2002}]{Johnston02}
  Johnston-Hollitt, M., Clay, R.W., Ekers, R.D., Wieringa, M.H.,
  \& Hunstead, R.W., 2002, in ``The Universe at low radio frequencies'', Pramesh
Rao, G. Swarup \& Gopal-Krishna (eds.), IAU Symp., 199, 157

\bibitem[\protect\citeauthoryear{Johnston-Hollitt}{2004}]{Johnston04}
  Johnston-Hollitt, M., 2004, in ``The riddle of cooling flows in
  galaxies and clusters of galaxies'', T. Reiprich, J. Kempner \&
  N. Soker (eds.), published electronically at
  http://www.astro.virginia.edu/coolflow/

\bibitem[\protect\citeauthoryear{Jones \& Forman}{1999}]{Jones99}
  Jones, C., \& Forman, W., \aspj{1999}{511}{65}

\bibitem[\protect\citeauthoryear{Kapferer et al.}{2006}]{Kapferer06}
  Kapferer, W., Ferrari, C., Domainko, W., et al., \aeta{2006}{447}{827}

\bibitem[\protect\citeauthoryear{Kassim et al.}{2001}]{Kassim01}
  Kassim, N.E., Clarke, T.E., En{\ss}lin, T.A., Cohen, A.S., \& Neumann,
  D.M., \aspj{2001}{559}{785}

\bibitem[\protect\citeauthoryear{Kempner \& Sarazin}{2001}]{Kempner01}
  Kempner, J.C., \& Sarazin, C.L., \aspj{2001}{548}{639}

\bibitem[\protect\citeauthoryear{Kempner et al.}{2004}]{Kempner04a} 
Kempner, J.C., Blanton, E.L., Clarke, T.E., et al., 2004, in ``The
riddle of cooling flows in galaxies and clusters of galaxies'',
T. Reiprich, J. Kempner \& N. Soker (eds.), published electronically
at http://www.astro.virginia.edu/coolflow/

\bibitem[\protect\citeauthoryear{Kempner \& David}{2004}]{Kempner04b} 
Kempner, J.C., \& David, L.P., \mnras{2004}{349}{385}

\bibitem[\protect\citeauthoryear{Kim et al.}{1990}]{Kim90}
  Kim, K.T., Kronberg, P.P., Dewdney, P.E., \& Landecker, T.L., 
\aspj{1990}{355}{29}

\bibitem[\protect\citeauthoryear{Kim et al.}{1991}]{Kim91}
  Kim, K.T., Tribble, P.C., \& Kronberg, P.P, \aspj{1991}{379}{80}

\bibitem[\protect\citeauthoryear{Large et al.}{1959}]{Large59}
Large, M.I., Mathewson, D.S., \& Haslam, C.G., 1959, Nature, 183, 1663

\bibitem[\protect\citeauthoryear{Lawler \& Dennison}{1982}]{Lawler82}
  Lawler, J.M., \& Dennison, B., \aspj{1982}{252}{81}

\bibitem[\protect\citeauthoryear{Liang et al.}{2000}]{Liang00}
  Liang, H., Hunstead, R.W., Birkinshaw, M., \& Andreani, P., 
\aspj{2000}{544}{686}

\bibitem[\protect\citeauthoryear{Longair}{1981}]{Longair81} Longair,
  M.S., 1981,  High energy astrophysics (Cambridge Univ.
  Press)

\bibitem[\protect\citeauthoryear{Markevitch et al.}{1999}]{Markevitch99}
Markevitch, M., Sarazin, C.L., \& Vikhlinin, A., \aspj{1999}{521}{526}

\bibitem[\protect\citeauthoryear{Markevitch et al.}{2000}]{Markevitch00}
  Markevitch, M., Ponman, T.J., Nulsen P.E.J., et al., \aspj{2000}{541}{542}

\bibitem[\protect\citeauthoryear{Markevitch \& Vikhlinin}{2001}]{Markevitch01} 
Markevitch, M., \& Vikhlinin, A., \aspj{2001}{563}{95}

\bibitem[\protect\citeauthoryear{Markevitch et al.}{2005}]{Markevitch05}
Markevitch, M., Govoni, F., Brunetti, G., \& Jerius, D., \aspj{2005}{627}{733}

\bibitem[\protect\citeauthoryear{Markevitch \& Vikhlinin}{2007}]{Markevitch07} 
Markevitch, M., \& Vikhlinin, A., 2007, Phys. Rep., 443, 1

\bibitem[\protect\citeauthoryear{Maurogordato et al.}{2007}]{Maurogordato07}
 Maurogordato, S., Cappi, A., Ferrari, C., et al., \aetap{2007}
 (astro-ph/0712.2715)

\bibitem[\protect\citeauthoryear{McNamara et al.}{2007}]{McNamara07}
McNamara, B.R., B\^irzan, L., Rafferty, D.A., et al., 2007, in
``Extragalactic jets: theory and observation from radio to gamma
ray'', T.A. Rector \& D.S. De Young (eds.), ASP Conference Series, in
press (astro-ph/0708.0579)

\bibitem[\protect\citeauthoryear{Meisenheimer et al.}{1989}]{Meisenheimer89}
Meisenheimer, K., Roser, H.-J., Hiltner, P.R., et al., \aeta{1989}{219}{63}

\bibitem[\protect\citeauthoryear{Miniati}{2003}]{Miniati03} 
Miniati, F., \mnras{2003}{342}{1009}

\bibitem[\protect\citeauthoryear{Mohr et al.}{1996}]{Mohr96} 
Mohr, J., Geller, M.J., \& Wegner, G., \aj{1996}{112}{1816}

\bibitem[\protect\citeauthoryear{Mulchaey}{1996}]{Mulchaey96} 
Mulchaey, J.S., Davis, D.S., Mushotzky, R.F., \& Burstein, D., 
\aspj{1996}{456}{80}

\bibitem[\protect\citeauthoryear{Murgia et al.}{2004}]{Murgia04}
Murgia M., Govoni F., Feretti L., et al., \aeta{2004}{424}{429}

\bibitem[\protect\citeauthoryear{Murgia et al.}{2005}]{Murgia05} 
Murgia, M., Parma, P., de Ruiter, H.R., et al., 2005, in ''X-ray and
radio connections'', L.O. Sjouwerman \& K.K Dyer (eds.), published
electronically by NRAO, http://www.aoc.nrao.edu/events/xraydio

\bibitem[\protect\citeauthoryear{Nakazawa et al.}{2007}]{Nakazawa07} 
Nakazawa, K., Makishima, K., \& Fukazawa, Y., 2007,  PASJ, 59, 167

\bibitem[\protect\citeauthoryear{Neumann \& Arnaud}{1999}]{Neumann99}
  Neumann, D.M., \& Arnaud, M., \aeta{1999}{348}{711}

\bibitem[\protect\citeauthoryear{Neumann \& Arnaud}{2001}]{Neumann01}
  Neumann, D.M., \& Arnaud, M., \aeta{2001}{373}{L33}

\bibitem[\protect\citeauthoryear{Newman et al.}{2002}]{Newman02} 
Newman, W.I., Newman, A.I., \& Rephaeli, Y., \aspj{2002}{575}{755}

\bibitem[\protect\citeauthoryear{Orr\`u et al.}{2007}]{Orru07}
  Orr\`u, E., Murgia, M., Feretti, L., et al., \aeta{2007}{467}{943}

\bibitem[\protect\citeauthoryear{Pacholczyk}{1970}]{Pacholczyk70}
  Pacholczyk, A.G., 1970,  Radio astrophysics, (Freeman Eds.)

\bibitem[\protect\citeauthoryear{Paerels et al.}{2008}]{Paerels08} 
Paerels, F., Kaastra, J., Ohashi, T., et al., 2008, SSR, in press

\bibitem[\protect\citeauthoryear{Parma et al.}{2007}]{Parma07}
Parma, P., Murgia, M., de Ruiter, et al., \aeta{2007}{470}{875}

\bibitem[\protect\citeauthoryear{Perley \& Taylor}{1991}]{Perley91} 
Perley R.A., \& Taylor G.B., \aj{1991}{101}{1623}

\bibitem[\protect\citeauthoryear{Petrosian}{2001}]{Petrosian01}
  Petrosian, V., \aspj{2001}{557}{560}

\bibitem[\protect\citeauthoryear{Petrosian}{2003}]{Petrosian03}
  Petrosian, V., 2003,  ASP Conf. Series, 301, 337

\bibitem[\protect\citeauthoryear{Petrosian et al.}{2008}]{Petrosian08}
  Petrosian, V., Rephaeli, Y., \& Bykov, A.M., 2008, SSR, in
  press

\bibitem[\protect\citeauthoryear{Pfrommer \&
    En{\ss}lin}{2004}]{Pfrommer04} Pfrommer, C., \& En{\ss}lin, T., 
\aeta{2004}{413}{17}

\bibitem[\protect\citeauthoryear{Poggianti et al.}{2004}]{Poggianti04}
Poggianti, B.M., Smail, I., Dressler, A., et al., \aspj{2004}{601}{197}

\bibitem[\protect\citeauthoryear{Rephaeli}{1977}]{Rephaeli77}
  Rephaeli, Y., \aspj{1977}{212}{608}

\bibitem[\protect\citeauthoryear{Rephaeli}{1979}]{Rephaeli79}
  Rephaeli, Y., \aspj{1979}{227}{364}

\bibitem[\protect\citeauthoryear{Rephaeli et al.}{1999}]{Rephaeli99}
  Rephaeli, Y., Gruber, D., \& Blanco, P., \aspj{1999}{511}{21}

\bibitem[\protect\citeauthoryear{Rephaeli \& Gruber}{2003}]{Rephaeli03}
  Rephaeli, Y., \& Gruber, D., \aspj{2003}{595}{137}

\bibitem[\protect\citeauthoryear{Rephaeli et al.}{2006}]{Rephaeli06}
  Rephaeli, Y., Gruber, D., \& Arieli, Y., \aspj{2006}{649}{673}

\bibitem[\protect\citeauthoryear{Rephaeli et al.}{2008}]{Rephaeli08}
  Rephaeli, Y., Nevalainen, J., Ohashi, T., \& Bykov, A.M., 2008, SSR, in
  press

\bibitem[\protect\citeauthoryear{Rizza et al.}{2000}]{Rizza00} 
Rizza, E., Loken, C., Bliton, M., Roettiger, K., \& Burns, J. O., 
\aj{2000}{119}{21}

\bibitem[\protect\citeauthoryear{R\"ottgering et al.}{1997}]{Roettgering97}
R\"ottgering, H.J.A., Wieringa, M.H., Hunstead, R.W., \& Ekers, R.D.,
\mnras{1997}{290}{577}

\bibitem[\protect\citeauthoryear{R\"ottgering}{2003}]{Roettgering03}
R\"ottgering, H.J.A., 2003,  New Astron. Rev., 47, 405

\bibitem[\protect\citeauthoryear{Rudnick \& Blundell}{2003a}]{Rudnick03a} 
Rudnick, L., Blundell, K.M., 2003a, in ``The riddle of
cooling flows in galaxies and clusters of galaxies'', T. Reiprich,
J. Kempner \& N. Soker (eds.), available electronically at 
http://www.astro.virginia.edu/coolflow/proc.php

\bibitem[\protect\citeauthoryear{Rudnick \& Blundell}{2003b}]{Rudnick03b}
Rudnick, L., \& Blundell, K.M., \aspj{2003b}{588}{143}

\bibitem[\protect\citeauthoryear{Sarazin \& Kempner}{2000}]{Sarazin00}
Sarazin, C.L., \& Kempner, J.C., \aspj{2000}{533}{73}

\bibitem[\protect\citeauthoryear{Sarazin}{2005}]{Sarazin05} 
Sarazin, C.L., 2005, in ``X-ray and radio connections'',
L.O. Sjouwerman \& K.K Dyer (eds.), published electronically by NRAO,
http://www.aoc.nrao.edu/events/xraydio

\bibitem[\protect\citeauthoryear{Sauvageot et al.}{2005}]{Sauvageot05}
Sauvageot, J.-L., Belsole, E., \& Pratt, G.W., \aeta{2005}{444}{673}

\bibitem[\protect\citeauthoryear{Schindler}{2002}]{Schindler02}
Schindler, S., 2002, in ``Merging processes in galaxy clusters'',
L. Feretti, I.M. Gioia \& G. Giovannini (eds.), Astrophys. Sp. Sc. Lib.
(Kluwer), 272, 229

\bibitem[\protect\citeauthoryear{Schuecker \& B\"ohringer}{1999}]{Schuecker99}
Schuecker, P., \& B\"ohringer, H., 1999, in ``Diffuse thermal and
relativistic plasma in galaxy clusters'', H. B\"ohringer, L. Feretti \&
P. Schuecker (eds.),  MPE Report, 271, 43

\bibitem[\protect\citeauthoryear{Sijbring}{1993}]{Sijbring93}
Sijbring, D., 1993, ``A radio continuum and H~I line study of
the Perseus cluster'', PhD Thesis, Groningen

\bibitem[\protect\citeauthoryear{Slee et al.}{2001}]{Slee01} Slee,
  O.B., Roy, A.L., Murgia, M., Andernach, H., \& Ehle, M., \aj{2001}{122}{1172}

\bibitem[\protect\citeauthoryear{Subrahmanyan}{2003}]{Subrahmanyan03} 
Subrahmanyan, R., Beasley, A.J., Goss, W.M., Golap, K., \& Hunstead,
R.W., \aj{2003}{125}{1095}

\bibitem[\protect\citeauthoryear{Sunyaev \& Zel'dovich}{1972}]{Sunyaev72}
Sunyaev, R.A., \& Zel'dovich, Y.B., 1972,  Comments on Astrophys.
  Space Phys., 4, 173

\bibitem[\protect\citeauthoryear{Taylor \& Perley}{1993}]{Taylor93}
Taylor, G.B., \& Perley R.A., \aspj{1993}{416}{554}

\bibitem[\protect\citeauthoryear{Taylor et al.}{2001}]{Taylor01}
Taylor, G.B., Govoni, F., Allen, S.W., \& Fabian, A.C., 
\mnras{2001}{326}{2}

\bibitem[\protect\citeauthoryear{Taylor et al.}{2002}]{Taylor02}
Taylor, G.B., Fabian, A.C., \& Allen, S.W., \mnras{2002}{334}{769}

\bibitem[\protect\citeauthoryear{Taylor et al.}{2007}]{Taylor07}
Taylor, G.B., Fabian, A.C., Gentile, G., et al., \mnras{2007}{382}{67}

\bibitem[\protect\citeauthoryear{Thierbach et al.}{2003}]{Thierbach03}
Thierbach, M., Klein, U., \& Wielebinski, R., \aeta{2003}{397}{53}

\bibitem[\protect\citeauthoryear{V\"olk et al.}{1996}]{Voelk96}
V\"olk, H., Aharonian, F.A., \& Breitschwerdt, D., \SpaceS{1996}{75}{279}

\bibitem[\protect\citeauthoryear{V\"olk \& Atoyan}{1999}]{Voelk99}
V\"olk, H., \& Atoyan, A.M., 1999,  Astrop. Phys., 11, 73

\bibitem[\protect\citeauthoryear{Vallee et al.}{1986}]{Vallee86}
Vall\'ee, J.P., MacLeod, M.J., \& Broten, N.W., \aeta{1986}{156}{386}

\bibitem[\protect\citeauthoryear{Vogt \& En{\ss}lin}{2003}]{Vogt03}
Vogt, C., \& En{\ss}lin, T.A., \aeta{2003}{412}{373}

\bibitem[\protect\citeauthoryear{Vogt \& En{\ss}lin}{2005}]{Vogt05}
Vogt, C., \& En{\ss}lin, T.A., \aeta{2005}{434}{67}

\bibitem[\protect\citeauthoryear{Venturi et al.}{2007}]{Venturi07}
  Venturi, T., Giacintucci, S., Brunetti, G., et al., \aeta{2007}{463}{937}

\bibitem[\protect\citeauthoryear{Vikhlinin et al.}{2001}]{Vikhlinin01}
Vikhlinin, A., Markevitch, M., \& Murray, S.S., \aspj{2001}{549}{L47}

\bibitem[\protect\citeauthoryear{Willson}{1970}]{Willson70}
Willson, M.A.G., \mnras{1970}{151}{1}


\end{thebibliography}
\end{document}